**Submitted to Magnetic Resonance in Medicine**

# *Improved unsupervised physics-informed deep learning for intravoxel incoherent motion modeling and evaluation in pancreatic cancer patients*

Misha P.T. Kaandorp[1,2,3,*], Sebastiano Barbieri[4], Remy Klaassen[5], Hanneke W.M. van Laarhoven[5], Hans Crezee[1], Peter T. While[2,3], Aart J. Nederveen[1], Oliver J. Gurney-Champion[1]

[1] Department of Radiology and Nuclear Medicine, Cancer Center Amsterdam, Amsterdam UMC, University of Amsterdam, Amsterdam, The Netherlands

[2] Department of Radiology and Nuclear Medicine, St. Olav's University Hospital, Trondheim, Trondheim, Norway

[3] Department of Circulation and Medical Imaging, NTNU – Norwegian University of Science and Technology, Trondheim, Norway

[4] Centre for Big Data Research in Health, UNSW, Sydney, Australia

[5] Department of Medical Oncology, Cancer Center Amsterdam, Amsterdam UMC, University of Amsterdam, Amsterdam, The Netherlands

[*] **Corresponding Author**

**Name:** Misha Pieter Thijs Kaandorp

**Department:** Radiology and Nuclear Medicine

**Institute:** St. Olav's University Hospital / NTNU – Norwegian University of Science and Technology

**Adress:** Ragnhilds gate 15, 7030 Trondheim, Norway

**Email:** mpkaando@stud.ntnu.no

### *Abstract*

**Purpose**: Earlier work showed that IVIM-NET$_{orig}$, an unsupervised physics-informed deep neural network, was faster and more accurate than other state-of-the-art intravoxel-incoherent motion (IVIM) fitting approaches to DWI. This study presents a substantially improved version: IVIM-NET$_{optim}$ and characterizes its superior performance in pancreatic cancer patients.

**Method:** In simulations (SNR=20), the accuracy, independence and consistency of IVIM-NET were evaluated for combinations of hyperparameters (fit *S0*, constraints, network architecture, # hidden layers, dropout, batch normalization, learning rate), by calculating the normalized root-mean-square error (NRMSE), Spearman's $\rho$, and the coefficient of variation (CV$_{NET}$), respectively. The best performing network, IVIM-NET$_{optim}$ was compared to least squares (LS) and a Bayesian approach at different SNRs. IVIM-NET$_{optim}$'s performance was evaluated in an independent dataset of twenty-three patients with pancreatic ductal adenocarcinoma (PDAC). Fourteen of the patients received no treatment between two repeated scan sessions and nine received chemoradiotherapy between the repeated sessions. Intersession within-subject standard deviations (wSD) and treatment-induced changes were assessed.

**Results**: In simulations (SNR=20), IVIM-NET$_{optim}$ outperformed IVIM-NET$_{orig}$ in accuracy (NRMSE(*D*)=0.177 vs 0.196; NMRSE(*f*)=0.220 vs 0.267; NMRSE(*D\**)=0.386 vs 0.393), independence ($\rho$(*D\**,*f*)=0.22 vs 0.74) and consistency (CV$_{NET}$(*D*)=0.013 vs 0.104; CV$_{NET}$(*f*)=0.020 vs 0.054; CV$_{NET}$(*D\**)=0.036 vs 0.110). IVIM-NET$_{optim}$ showed superior performance to the LS and Bayesian approaches at SNRs<50. In vivo, IVIM-NET$_{optim}$ showed significantly less noisy parameter maps with lower wSD for *D* and *f* than the alternatives. In the treated cohort, IVIM-NET$_{optim}$ detected the most individual patients with significant parameter changes compared to day-to-day variations.

**Conclusion:** IVIM-NET$_{optim}$ is recommended for accurate, informative and consistent IVIM fitting to DWI data.

## *1. Introduction*

The intravoxel incoherent motion (IVIM) model (1) for diffusion-weighted imaging (DWI) shows great potential for estimating predictive and prognostic cancer imaging biomarkers (2–5). In the IVIM model, DWI signal is described by a bi-exponential decay, of which one component is attributed to conventional molecular diffusion and the other to the incoherent bulk motion of water molecules, typically credited to capillary blood flow. Hence, IVIM simultaneously provides information on diffusion ($D$; diffusion coefficient), capillary microcirculation ($D^*$; pseudo-diffusion coefficient) and the perfusion fraction ($f$) without the use of a contrast agent (6–8). However, despite IVIM's great potential (2–5), it is rarely used clinically. Two major hurdles preventing routine clinical use of IVIM are its poor image quality and the long fitting time (9–11). Tackling these shortcomings will help towards wider use of IVIM (12).

Currently, IVIM is often fitted using the conventional least squares (LS) algorithm. However, more accurate alternative approaches have been suggested (9). Until recently, Bayesian algorithms for IVIM fitting (13) were most promising regarding inter-subject variability (9), precision, accuracy (14), and smooth parameter maps, suggesting less noise (15). Conversely, Bayesian approaches are substantially slower ($9 \times 10^{-2}$ s/vox (11)) than the already slow LS approach ($8 \times 10^{-3}$ s/vox (11)). Furthermore, Bayesian approaches may lead to biased perfusion estimates of the IVIM model (16).

Recently, a promising alternative for IVIM fitting was introduced: estimating IVIM parameters with deep neural networks (DNNs). Initially, Bertleff et al. (17) introduced a supervised DNN for IVIM parameter estimation, in which the network was trained on simulated data for which the underlying parameters were known. However, the strong assumption of simulated training and test data being identically distributed could limit the network's performance in vivo, where noise behaves less ordered. We solved this shortcoming in earlier work (11), where we used unsupervised physics-informed deep neural networks (PI-DNNs) (18,19). PI-DNNs formulate a physics-informed-loss-function that finds learned parameters through an iterative process. In this case, the PI-DNN used consistency between the predicted signal from the IVIM model and the measured signal as a loss term in the DNN. This resulted in an unsupervised PI-DNN capable of training directly on patient data with no ground truth: IVIM-NET$_{orig}$. We demonstrated in both simulations and patient analysis that IVIM-NET$_{orig}$ is superior to the conventional LS approach and even performs (marginally) better than the Bayesian approach. Furthermore, IVIM-NET$_{orig}$'s fitting times were substantially lower ($4 \times 10^{-6}$ s/vox (11)) than the LS and Bayesian approaches. However, that proof of principle IVIM-NET study did not explore many hyperparameters and focused on volunteer data.

In this work, we hypothesize that IVIM-NET$_{orig}$ can be improved by exploring the architecture of the network, its training features and other hyperparameters. Hence, we characterized the performance of IVIM-NET for different hyperparameter settings by assessing the accuracy, independence and consistency of the estimated IVIM parameters in simulated IVIM data. Finally, we compared the performance of our optimized IVIM-NET to the LS approach and a Bayesian approach in patients with pancreatic ductal adenocarcinoma (PDAC) receiving neoadjuvant chemoradiotherapy (CRT) in terms of image quality, parameter to noise ratio, test-retest reproducibility and sensitivity to treatment effects.

## 2. Methods

### 2.1 IVIM-NET

We initially implemented the original PI-DNN (IVIM-NET$_{orig}$) (11) in Python 3.8 using PyTorch 0.4.1 (20). The input layer consisted of neurons that took the normalized DWI signal *S(b)/S(b=0)* as input, where *S(b)* is the measured signal at diffusion weighting *b* (*b* value). The input layer was followed by three fully connected hidden layers. Each hidden layer had several neurons equal to the number of measurements (*b* values and the number of repeated measures) and each neuron, in turn, contained an exponential linear unit activation function (21). The output layer of the network consisted of the three IVIM parameters (*D*, *f*, *D\**). To enforce the output layer to predict these IVIM parameters, two steps were taken. First, the absolute activation function was taken of the neuron's output (*X*) to constrain the predicted parameters, e.g. to compute *D*:

$$D = |X[1]| \qquad [1].$$

Second, a physics-based loss function was introduced that computed the mean squared error between the measured input signal, *S(b)*, and the predicted IVIM signal $S_{net}(b)$, which was obtained by inserting the predicted output parameters into the normalized IVIM model. Hence:

$$L = \frac{1}{|B|} \sum_{b \in B} \left( \frac{S(b)}{S(b=0)} - S_{net}(b) \right)^2 \qquad [2],$$

with

$$S_{net}(b) = f e^{-bD^*} + (1-f) e^{-bD} \qquad [3],$$

*w*here *B* is the total number of image acquisitions.

Next, we evaluated whether seven novel hyperparameters (Table 1; Figure 1) of IVIM-NET improved fitting results. First, instead of fixing *S0*, we added *S0* as an additional output parameter, to allow the system to correct for noise in *S(b=0)*. Second, to restrict parameter values to physiologically plausible ranges, scaled sigmoid activation functions instead of absolute activation functions were used to constrain the predicted parameters (Table 1), e.g. to compute *D*:

$$D = D_{min} + sigmoid(X[1]) * (D_{max} - D_{min}) \qquad [4],$$

where $D_{min}$ and $D_{max}$ are the fit boundaries. Third, we varied the number of hidden layers between 1 and 9. Fourth, we used dropout regularization (22) in all hidden layers except for the last one. Dropout randomly removes a set percentage of network-weights each iteration during training. Fifth, we used batch normalization (23) which normalizes the input by re-centering and re-scaling, and, consequently, preserves the representation ability of the network. Sixth, to reduce unwanted correlation between estimated parameter values, we implemented an alternative network architecture in which parameter values were predicted, in parallel, by independent sub-networks (Table 1; Figure 1). Furthermore, we evaluated different learning rates (LR) of the Adam optimizer (24), ranging from $1 \times 10^{-5}$ to $3 \times 10^{-2}$, and with constant $\beta=(0.9, 0.999)$.

In traditional deep learning, training and evaluation are done on separate datasets, but as this is an unsupervised DNN approach, training was done on the same data as evaluation (11,25). So, for simulations, these were simulated data, and in vivo, these were in vivo data. 90% of the data was used for training and 10% of the data was used for validation. Early stopping occurred when the validation loss did not improve over 10 consecutive training epochs. Given the large amount of training data and the limited number of network parameters, each epoch consisted of only 500 random batches. So, effectively the network saw $500 \times 128$ IVIM curves in between validations.

### 2.2 Simulations: characterization and optimization

100,000 IVIM curves were simulated to investigate the effects of different hyperparameters on the accuracy, independence and consistency of the estimated IVIM parameters. DWI signals were simulated based on Eq. 3 with $S0 = 1$, 11 $b$ values ($b = 0, 5, 10, 20, 30, 40, 60, 150, 300, 500,$ and $700$ s/mm$^2$), and pseudorandom uniformly sampled values of $D$: $0.5 \times 10^{-3}$ to $3 \times 10^{-3}$ mm$^2$/s, $f$: 5 to 55%, and $D^*$: $10 \times 10^{-3}$ to $100 \times 10^{-3}$ mm$^2$/s. These ranges were slightly broader than the typical values found in abdominal IVIM (26). Random Rician noise in the form of complex Gaussian noise was added to the curves with predefined SNR levels (constant noise amplitude over $b$ values; SNR defined at $b = 0$ as $S(b=0)/\sigma$, with $\sigma$ the standard deviation of the Gaussians) (27).

Accuracy was assessed as the normalized root-mean-square error (NRMSE) between the ground truth parameter values and the estimated IVIM parameters.

Independence of the parameter estimates was assessed by the Spearman rank correlation coefficients ($\rho$) between all parameter pairs. As the simulated data were independent and random, a $\rho$ should be 0. The absolute value of $\rho$ was taken, as both positive and negative deviations from zero are equally undesirable. Some networks always returned the same value for $D^*$, independent of the input data (Supporting Information Figure S1). For such cases, $\rho$ is technically undefined. As these cases are undesirable $\rho$ was set to 1.

As training a DNN is a stochastic process, training on the same dataset results in different final network-weights, and consequently, different predictions on the same data. To assess the consistency of estimated parameter values, each network variant was trained 50 times on identical data, where each repeat had a new random initialization, dropout and batch selection. The normalized coefficient of variation per parameter over the repeated trainings ($CV_{NET}$) was taken as a measure of the consistency:

$$CV_{NET} = \frac{1}{\bar{x}_{true}} \sqrt{\frac{1}{n \times (m-1)} \sum_{i=1}^{n} \sum_{j=1}^{m} (x_{i,j} - \bar{x}_i)^2} \qquad [5],$$

where $\bar{x}_{true}$ is the mean simulated IVIM parameter value, $n$ the number of simulated curves, $m$ the number of repeated trainings, $x_{i,j}$ is the $j$th repeated prediction of the $i$th simulated decay curve, $\bar{x}_i$ is the mean over the repeated $m$ predictions of the $i$th simulated signal curve. As the LS and Bayesian approaches are deterministic, their $CV_{NET}$ was zero.

As a result of the repeated training, we obtained 50 values for the NRMSEs and $\rho$'s. Therefore, the median and interquartile ranges were reported.

As a baseline for comparison, we evaluated the IVIM parameters ($D$, $f$, $D^*$) in IVIM-NET$_{orig}$, the LS and Bayesian approaches. We used the Levenberg-Marquardt non-linear algorithm for the LS fit (28,29). For the Bayesian approach, we used the algorithm from previous work (11). For both the LS and Bayesian approaches $S0$ was included as a fit parameter. The Bayesian approach used a data-driven lognormal prior for $D$ and $D^*$, and a beta distribution for $f$ and $S0$. The prior distributions were determined empirically by fitting these distributions to the results from the LS approach on the same dataset. The maximum a posteriori probability was used as an estimate of the IVIM parameters. The LS and Bayesian approaches were performed with fit boundaries of $D$: $0 \times 10^{-3}$ to $5 \times 10^{-3}$ mm$^2$/s, $f$: 0 to 70%, $D^*$: $5 \times 10^{-3}$ to $300 \times 10^{-3}$ mm$^2$/s, and $S0$: 0.7 to 1.3.

After baseline characterization, IVIM-NET was optimized by testing various combinations of the hyperparameters (Table 1; Figure 1). Previous studies reported reliable SNR values of IVIM in the abdomen between 10 and 40 (30–33). So, to simulate reliable abdominal IVIM signals, an SNR of 20 was chosen for hyperparameter evaluation. We trained the network on the simulated signals using every combination of the following options: fit $S0$ parameters, absolute or sigmoid constraints, parallel network, dropout and batch normalization - while fixing the number of hidden layers to 3 (used in IVIM-NET$_{orig}$, Table 1) and the LR to $1 \times 10^{-4}$. Bound intervals of the sigmoid activation functions were chosen 60% wider (30% at each side) than the fit boundaries of the LS and Bayesian approaches to compensate for decreasing gradients at the asymptotes of the sigmoid function. In an exploratory phase, we found that reducing the LR from $1 \times 10^{-3}$ (IVIM-NET$_{orig}$) to $1 \times 10^{-4}$ was essential for obtaining networks with improvements in accuracy, independence and consistency. Each network (i.e. combination of hyperparameters) received a ranking in each of the 9 performance measures (3 metrics for 3 parameters), and these 9 ranks were summed. The best performing network was then chosen by selecting the network with the lowest summed rank.

With the best options for the fit *S0* parameters, constraints, parallel network, dropout and batch normalization, we tested the performance of the network as a function of the LR and the number of hidden layers (Table 1). From those results, we finally selected the best performing optimized network by again selecting the lowest summed rank: IVIM-NET$_{optim}$. IVIM-NET$_{optim}$'s performance was then characterized and compared to the LS approach, Bayesian approach and IVIM-NET$_{orig}$ for SNR values between 8 (low) and 100 (high).

### 2.3 Verification in patients with PDAC

We used two IVIM datasets of patients with PDAC to validate IVIM-NET$_{optim}$'s performance in vivo: one dataset to assess test-retest reproducibility, and one to test whether we can detect treatment effects. Both studies were approved by our local medical ethics committee and all patients gave written informed consent.

Both datasets (NCT01995240; NCT01989000) were published earlier (9,34,35). The first dataset consists of 14 patients with locally advanced or metastatic PDAC who underwent IVIM in two separate imaging sessions (average 4.5 days apart, range: 1–8 days) with no treatment in-between. The second dataset consisted of 9 PDAC patients with (borderline) resectable PDAC who received CRT as part of the PREOPANC study (36) where patients were scanned before and after CRT.

MRI data were acquired using a 3T MRI scanner (Ingenia, Philips, Best, The Netherlands). A respiratory triggered (navigator on liver dome) 2D multi-slice diffusion-weighted echo-planar imaging was used with parameters: TR > 2200 ms (depending on respiration speed), TE = 45 ms, flip angle = 90 deg, FOV = 432 x 108 mm$^2$, acquisition matrix = 144 × 34, 18 slices, slice thickness = 3.7 mm and 12 *b* values (directions): 0 (15), 10 (9), 20 (9), 30 (9), 40 (9), 50 (9), 75 (4), 100 (12), 150 (4), 250 (4), 400 (4) and 600 (16) mm$^2$/s. Fat suppression was carried out with a gradient reversal during slice selection and spectral presaturation with inversion recovery. Diffusion gradient times were 10.1 ms with a delay between diffusion gradients onset of 22.6 ms.

DWI images were co-registered to a reference volume consisting of a mean DWI image over all *b* values using deformable image registration in Elastix (37). A radiologist (10 years' experience in abdominal radiology) and researcher (4 years' experience in contouring pancreatic cancer) drew a region of interest (ROI) in the tumor in consensus. IVIM parameter maps of *D*, *f* and *D\** were derived using the LS approach, Bayesian approach and IVIM-NET$_{optim}$. Background voxels were removed automatically before fitting by removing voxels with *S(b=0)<0.5×median(S(b=0))*. Fitting was done without averaging over the diffusion directions. IVIM-NET$_{optim}$ was trained on all combined patient data. Values under 0 for *D*, *f* and *D\** were considered not physiologic and set to 0, and for further statistics, values of $D*$ were masked where $f < 5\%$ as $D*$ only "exists" in perfused voxels. All computations were carried out on a single core of a conventional desktop computer (CPU: Intel Core i7-8700 CPU at 3.20 GHz). The average fitting time of each algorithm was recorded.

Firstly, the parameter maps were compared qualitatively in terms of feature clarity, and by visually assessing consistency of fit to the IVIM signal in pairs of neighboring voxels. For a quantitative comparison, the parameter to noise ratio (*PNR*) of the parameter maps was estimated in a homogeneous 2D ROI (>20 voxels) in the liver. *PNR* was defined as *mean/STD* of the homogenous ROIs and was calculated for each scan separately. We tested whether the difference in *PNR* between fit approaches was significant using paired t-tests.

To determine clinical usefulness of IVIM-NET, we investigated whether we could detect changes in parameter values throughout CRT by comparing patients receiving treatment to the baseline test-retest repeatability. This analysis was performed with the median parameter values from within the ROIs. To evaluate test-retest repeatability, intersession within-subject standard deviation (wSD) (38) was calculated for each IVIM parameter using the data from the patients with repeated baseline scans. Bland-Altman plots were plotted for patients from both cohorts. We calculated the 95% confidence intervals (95CIs) from the patients with repeated scans at baseline (assuming zero offsets). In the treated cohort, we used a paired t-test to test whether parameters had significantly changed due to treatment within the cohort (significance level $\alpha=0.05$). Furthermore, patients from the treatment cohort were added to the Bland-Altman plots and individual patients who had changes exceeding the 95CIs were considered to have significant changes in tumor microstructure (39).

## 3. Results

### 3.1 Simulations: characterization and optimization

The original network, IVIM-NET$_{orig}$, showed substantially lower NRMSE for all estimated parameters than the LS and Bayesian approaches. However, IVIM-NET$_{orig}$ had strong correlations between $D*$ and $f$ (high $\rho(D*,f)$; Table 2 and Figure 2D), and had considerable CV$_{NET}$.

The NRMSE, $\rho$ and CV$_{NET}$ for all hyperparameter combinations are shown in the Supporting Information Figures S2–S9. The summarizing sum of ranks (Supporting Information Figures S5 and S9) allowed us to determine IVIM-NET$_{optim}$ (Table 1). IVIM-NET$_{optim}$ resolved the high $\rho(D*,f)$ found in IVIM-NET$_{orig}$ (Table 2; Figure 2D, 2E) and substantially reduced the NRMSE and CV$_{NET}$. Single changes away from IVIM-NET$_{optim}$ can lead to marginally better NRMSE, lower $\rho$ or lower CV$_{NET}$ (Figure 3), but only at a cost to the other two attributes. It is clear that the reduced $\rho(D*,f)$ cannot be attributed to a single parameter, but was a result of the combination of sigmoid constraints and batch normalization (Supporting Information Figure S3). Adding dropout (10%), fitting $S0$ and using our parallel network design decreased the NRMSE, while still having a low $\rho(D*,f)$ (Table 2). Increasing dropout in IVIM-NET$_{optim}$ or using a single network architecture resulted in similar NRMSE, however, increased $\rho(D^*,f)$ (Figure 3; Supporting Information Figures S2-S4). Generally, increasing the number of hidden layers resulted in a marginally higher $\rho$, and lower NRMSE and CV$_{NET}$. A too high/low LR (Supporting Information Figures S5–S7) caused higher NRMSEs and less consistency.

IVIM-NET$_{optim}$ was superior to the LS and Bayesian approaches for SNRs 8–33. Compared to IVIM-NET$_{orig}$, IVIM-NET$_{optim}$ was associated with improved NRMSE for $f$ and $D$ at all SNRs (Figure 4). For $D^*$, the networks performed similarly regarding NRMSE, with IVIM-NET$_{optim}$ performing slightly better at SNRs > 20 and IVIM-NET$_{orig}$ for SNRs < 20. IVIM-NET$_{optim}$ had lower $\rho(D^*,f)$ than IVIM-NET$_{orig}$ and improved CV$_{NET}$ for all SNR levels.

### 3.2 Verification in patients with PDAC

Examples of parameter maps computed with the LS approach, Bayesian approach and IVIM-NET$_{optim}$ together with two individual voxel fits of PDAC patients from both cohorts are presented in Figures 5 and 6. Additional parameter maps of 10 other PDAC patients are shown in the Supporting Information Figures S11–S20. Qualitatively comparing the parameter maps shows that IVIM-NET$_{optim}$ has very similar voxel values as the LS and Bayesian approaches for most voxels. However, where the LS and Bayesian approaches sometimes show "noisy" voxels (i.e. different from their neighbors) with substantially higher $f$ (order of 1) and lower $D$ (some voxels as low as $D = 0$ mm$^2$/s; Supporting Information Figures S11, S12) and $D^*$ (oftentimes to the lower bound of $D^*$; Supporting Information Figures S12, S14–16, S19), IVIM-NET$_{optim}$ often sticks to sensible $D$, $f$ and $D^*$ that are similar to the neighboring voxels resulting in more homogenous parameter maps. Note that IVIM-NET is fitted at a per-voxel level and is unaware of the voxel location. Quantitatively evaluating the parameter maps shows that IVIM-NET$_{optim}$ had significantly better *PNR* than both the LS and Bayesian approaches for $D$ and $D^*$ (Table 3) and significantly better *PNR* than the LS approach for $f$.

In the test-retest cohort, IVIM-NET$_{optim}$ showed the lowest wSD for $D$ and $f$ (Table 3), while the Bayesian approach had the lowest wSD for $D^*$. When averaging IVIM parameters for the repeated patient scans, IVIM-NET$_{optim}$ computed a higher $D$, lower $f$ and higher $D^*$ than the LS and Bayesian approaches (Table 3). The repeated scans are visualized as black x's in the Bland-Altman plots, together with their 95CIs in Figure 7.

When considering the CRT patients as a whole, IVIM-NET$_{optim}$ found a significant increase in mean $D$ and $f$ after treatment, whereas the LS approach found only a significant increase in $D$ after treatment (Table 3). The Bayesian approach found no significant change in IVIM parameters.

Figure 7 shows the individual change in IVIM parameter values of patients receiving CRT compared to the 95CIs of the test-retest cohort. With 10 significant changes, IVIM-NET$_{optim}$ detected the most patients with significant parameter changes after CRT, with 4 individual patients with increased $D$, 3 patients with increased $f$ and 3 patients with changes in $D*$. In comparison, the LS and Bayesian approaches detected only 2 and 3 significant parameter changes, respectively.

The average interference fitting time of IVIM-NET$_{optim}$ after training was $3.0 \times 10^{-5}$ s/vox, whereas the average fitting times of the LS and Bayesian approaches were $4.2 \times 10^{-3}$ s and $1.0 \times 10^{-1}$ s/vox, respectively. The median training time for IVIM-NET$_{optim}$ (20 repeats) was 572 s with a range of 401 to 685 s, which for our dataset resulted in $2.9 \times 10^{-4}$ s/vox training.

## 4. Discussion

This study is the first to show the potential clinical benefit of DNNs for IVIM fitting to DWI data in a patient cohort. We successfully developed and trained IVIM-NET$_{optim}$, an unsupervised PI-DNN IVIM fitting approach to DWI data that predicts accurate, independent and consistent IVIM parameters in simulations and in vivo, in patients with PDAC. IVIM-NET$_{optim}$ consisted of a parallel network architecture with 2 hidden layers, batch normalization, dropout of 10%, sigmoid constraints and fitted $S0$. Optimized training was performed using an Adam optimizer with a LR of $3 \times 10^{-5}$. In simulations, IVIM-NET$_{optim}$ outperformed the original version, IVIM-NET$_{orig}$, by offering more accurate estimates of $D$, $f$ and $D*$, with substantially less correlation between the estimated parameters $D*$ and $f$ and more consistent parameter prediction. Furthermore, simulations demonstrated that IVIM-NET$_{optim}$ had substantially better accuracy than the conventional LS and state-of-the-art Bayesian approaches. Finally, in patients with PDAC, IVIM-NET$_{optim}$ also outperformed the alternatives. IVIM-NET$_{optim}$ showed the most detailed and significantly less noisy parameter maps, and a significant change in diffusion and perfusion fraction for the whole cohort receiving CRT. Furthermore, IVIM-NET$_{optim}$ was associated with the best test-retest repeatability (smallest wSD) for $D$ and $f$, which allowed it to detect the most patients with significant changes in all IVIM parameters after CRT.

IVIM-NET$_{optim}$ detected a significant positive trend in $D$ and $f$ for the whole cohort of patients receiving CRT, whereas the LS approach only found a significant positive trend in $D$. Also, IVIM-NET$_{optim}$ detected 4 patients with a significant parameter increase for $D$, whereas the LS approach only detected 1 patient. These findings strongly suggest IVIM-NET$_{optim}$ as a good alternative for IVIM fitting in PDAC patients. Findings from other studies support this increase in $D$ (40) and $f$ (41) during CRT in PDAC patients. In general, PDACs tend to have lower diffusion due to the impeded water movement of compressing cells (42). Furthermore, PDACs are typically hypoperfused, due to significant tumor sclerosis creating elevated interstitial pressure, which compresses tumor feeding vessels (41,43). Effective treatment leads to necrosis, which in turn leads to lower cell densities and reduced interstitial pressure, and consequently increased diffusion (44,45) and perfusion (43,46). Not all patients demonstrated a significant change induced by treatment. Therefore, using IVIM to discriminate between individual treatment effects may be feasible in the future. As treatment of these patients was part of induction therapy and patients received surgery directly after, overall survival cannot be attributed purely to CRT effects. Hence, given the limited number of patients and the diluted treatment effect, we did not compare overall survival between patients that showed potential treatment effects and others.

Our previous work (35) showed that the LS approach for IVIM fitting was sensitive to individual treatment effects. However, the high wSD limited the study to detect individual treatment effects. Furthermore, this work (35) used denoised DWI b-images that substantially degraded image sharpness and tumor boundaries were harder to detect (e.g. compare figures from this work to example figures from (9)). Conversely, our present study demonstrates that DNNs can estimate parameter maps directly from the noisy data resulting in sharp high-quality IVIM parameter maps.

For most voxels, IVIM-NET$_{optim}$ produces very similar estimates to the LS and Bayesian approaches. However, within the tumor ROI, IVIM-NET$_{optim}$ shows consistently different mean baseline parameters than the LS and Bayesian approaches (Table 3). We believe that there are two major contributors to this discrepancy in mean values. 1) The LS and Bayesian approaches have more noisy parameter maps with some individual voxels showing extreme estimates. 2) IVIM-NET$_{optim}$ is seemingly better at estimating parameters in regions of poor perfusion. The first observation is demonstrated by the individual voxel fits (Figures 5 and 6; Supporting Information Figures S11–S20) where the LS and Bayesian approaches occasionally compute noisy IVIM parameters with a substantially higher $f$, and lower $D$ and $D*$ than IVIM-NET$_{optim}$. As the LS approach minimizes the sum of the residuals, this parameter combination could describe better the noisy data. However, inspecting neighboring voxels with respectively similar noisy data show that the LS and Bayesian approaches are inconsistent in producing the same IVIM parameters, whereas IVIM-NET$_{optim}$ is more consistent. The second note is especially interesting for PDACs which are generally hypoperfused (41,43). Other studies reported an overestimation of perfusion parameters in poorly perfused tissue (47–49) and indeed, the LS and Bayesian approaches show high and noisy perfusion fractions maps in the PDACs. Conversely, IVIM-NET$_{optim}$ shows consistently low and less noisy perfusion in these regions. Another interesting observation is that in Figure 6, IVIM-NET$_{optim}$ shows a very similar $f$ map to the LS approach for almost all tissue, suggesting again that there is no bias. Yet, contrary to the LS approach, IVIM-NET$_{optim}$ has a homogeneously low perfusion map in the PDAC (i.e. Figures 5 and 6; Supporting information S11–S20). In the absence of ground truth in patients, we must rely on visual assessment of parameter map quality, the ability to detect treatment response, and the simulation results, in order to infer the best performing estimator. Given the evidence provided in this study, we argue that combined these factors rule in favor of IVIM-NET$_{optim}$.

Although IVIM-NET showed consistently better results both in simulations and in vivo, IVIM-NET predicts different IVIM parameters in repeated training. This causes a new sort of variability that, until now, was not an issue in fitting parameter maps. There may be methods to mitigate this variability. First, when probing treatment response, we would advise using one network such that this additional effect is not different pre- and post-treatment. Second, to reduce the variation, one could consider taking the median prediction from 10 repeated trainings instead. We did so in an exploratory study where we formed 5 groups of 10 networks and showed that the median of 10 networks was substantially more consistent, with $CV_{NET}$ values of $3.2 \times 10^{-3}$, $4.9 \times 10^{-3}$ and $9.3 \times 10^{-3}$ for $D$, $f$ and $D*$, respectively. Having a set of networks will also allow the user to estimate the variation on the predicted parameter. Finally, although we see this additional uncertainty, we would like to stress that it is secondary to the overall error of the LS approach, which is apparent from the fact that in the simulations, all 50 instances of IVIM-NET$_{optim}$ had lower NRMSE than the LS approach.

IVIM-NET$_{optim}$ outperformed IVIM-NET$_{orig}$ at SNRs 8–100 and was superior to the LS and Bayesian approaches for SNRs 8–33 (Figure 4). However, at extremely high SNR (SNR = 100; Figure 4), the LS approach outperformed IVIM-NET. The Levenberg-Marquardt algorithm for the LS function is an iterative function that finds a minimum of the squared difference. For a relatively smooth loss landscape and high SNR signal, the LS algorithm is designed to find the correct parameter estimates. However, at low SNR, the LS approach has trouble finding the correct parameters. This occurs either because the loss landscape is no longer smooth and hence it gets stuck in a local minimum, or, what we believe is more common, the noise has changed the signal such that the global optimum no longer is nearby the ground truth parameters. On the other hand, a DNN consists of a complex system that needs to encompass estimating the IVIM parameters for all voxels. It turns out that having been trained on all voxels enables better estimates for individual voxels at low SNR. We expect that DNNs focus on more consistent minima with parameter values that are more frequently observed. This might be similar to data-driven Bayesian fitting approaches (15,50). Conversely, IVIM-NET seems to reach a maximum accuracy at high SNR. Potentially more complex DNNs that are optimized with simulations done at high SNR could handle the subtle signal changes of the IVIM parameters at these SNRs. However, typical SNR values for IVIM data are < 50. Therefore, our findings suggest that using IVIM-NET instead of the LS and Bayesian approaches for IVIM fitting would be beneficial in a clinical setting.

The choice of the hyperparameters for IVIM-NET$_{optim}$ was based on an optimal combination of accuracy, independence and consistency across all IVIM parameters. However, other hyperparameter options may be more appropriate when characterizing an individual IVIM parameter (e.g. when an observer is only interested in $D$ and IVIM is only used to correct for perfusion). Figures S2–S9 of the Supporting Information can help interested readers select the best network for their purposes.

The high dependency between $D*$ and $f$ that appears in IVIM-NET$_{orig}$ could not be attributed to a single cause. Initially, we expected that this dependency originated in the fully connected shared hidden layers of the original network. However, $\rho$ remains substantial when adding the 'parallel network architecture' to IVIM-NET$_{orig}$ (Supporting Figure; Figure S10). Using IVIM-NET$_{optim}$ and a single network architecture showed slightly worse performance in simulations regarding $\rho(D*,f)$, but still had sufficient accuracy and consistency. The dependencies between the estimated IVIM parameters are not per se specific to unsupervised DNNs. For instance, similar dependencies between $D*$ and $D$ or $f$ were found in a different data-driven Bayesian fitting approach (9). For IVIM-NET$_{optim}$ these dependencies were small at clinical SNR values and similar to those of the LS approach.

Although simulation studies in parameter estimation are extremely valuable as the underlying parameter values are known, they also come with limitations. One limitation is that the noise characterization of real data can be diverse and hard to model. For instance, DWI artifacts caused by motion are not considered in simulations and may affect the results of fitting the IVIM model (51). Another limitation is the underlying assumption that data are perfectly bi-exponential. In reality, the IVIM model is a simplification and real data will be more complex.

## 5. Conclusion

We substantially improved the accuracy, independence and consistency of both diffusion and perfusion parameters from IVIM-NET by changing the network architecture and tuning hyperparameters. Our new IVIM-NET$_{optim}$ is considerably faster, and computes less noisy and more detailed parameter maps with substantially better test-retest repeatability for $D$ and $f$ than alternative state-of-the-art fitting methods. Furthermore, IVIM-NET$_{optim}$ was able to detect the most individual patients with significant changes in the IVIM parameters throughout CRT. These results strongly suggest using IVIM-NET$_{optim}$ for detection of treatment response in individual patients.

## Data availability statement

To stimulate a wider clinical implementation of IVIM, we have made the code of IVIM-NET available on GitHub: https://github.com/oliverchampion/IVIMNET. This includes some simple introductory examples in simulations and additional volunteer data. We encourage our peers to use it in their research.

## *Figures and Tables*

**Table 1:** Hyperparameter settings for training IVIM-NET, including the settings for IVIM-NET$_{orig}$ and IVIM-NET$_{optim}$.

| Hyperparameter | Values | IVIM-NET$_{orig}$ | IVIM-NET$_{optim}$ |
|---|---|---|---|
| Fit *S0* | True, False | False | True |
| Constraints | Sigmoid, Absolute | Absolute | Sigmoid |
| Parallel networks | True, False | False | True |
| Number of hidden layers | 1,2 3, 4, 5, 6, 7, 8, 9 | 3 | 2 |
| Dropout regularization | 0%, 10%, 20%, 30% | 0% | 10% |
| Batch normalization | True, False | False | True |
| Learning rate | $1 \times 10^{-5}$, $3 \times 10^{-5}$, $1 \times 10^{-4}$, $3 \times 10^{-4}$, $1 \times 10^{-3}$, $3 \times 10^{-3}$, $1 \times 10^{-2}$, $3 \times 10^{-2}$ | $1 \times 10^{-3}$ | $3 \times 10^{-5}$ |

**Table 2:** Normalized root-mean-square error (NRMSE), Spearman rank correlation coefficient ($\rho$) and normalized coefficient of variation ($CV_{NET}$) of the LS approach, Bayesian approach, IVIM-NET$_{orig}$ and IVIM-NET$_{optim}$ for the estimated parameters IVIM ($D$, $f$, $D*$) in simulations at SNR 20 for 50 repeated trainings. Values of IVIM-NET: median (interquartile range).

| | Least Squares | Bayesian | IVIM-NET$_{orig}$ | IVIM-NET$_{optim}$ |
|---|---|---|---|---|
| **NRMSE $D$ [fraction]** | 0.279 | 0.233 | 0.196 (0.190–0.214) | 0.177 (0.176–0.178) |
| **NRMSE $f$ [fraction]** | 0.387 | 0.281 | 0.267 (0.259–0.273) | 0.220 (0.218–0.222) |
| **NRMSE $D*$ [fraction]** | 0.805 | 0.575 | 0.393 (0.382–0.414) | 0.386 (0.381–0.390) |
| $\rho(D,D*)$ | 0.24 | 0.08 | 0.23 (0.17–0.28) | 0.20 (0.19–0.21) |
| $\rho(D,f)$ | 0.18 | 0.03 | 0.04 (0.02–0.09) | 0.01 (0.00–0.01) |
| $\rho(D*,f)$ | 0.20 | 0.13 | 0.74 (0.64–0.80) | 0.22 (0.23–0.2) |
| $CV_{NET}$ $D$ [fraction] | 0 | 0 | 0.104 | 0.013 |
| $CV_{NET}$ $f$ [fraction] | 0 | 0 | 0.054 | 0.020 |
| $CV_{NET}$ $D*$ [fraction] | 0 | 0 | 0.110 | 0.036 |

**Table 3:** Parameter to noise ratio (*PNR*) of homogenous liver tissue (top panel), intersession within-subject standard deviation (wSD) (middle panel), and mean IVIM parameters for the patients with treatment (bottom panel).

| *PNR* | *D* | *f* | *D\** |
|---|---|---|---|
| **Least squares** | 5.6 | 3.1 | 1.8 |
| **Bayesian** | 6.3 | 4.0 | 2.3 |
| **IVIM-NET**optim | **8.1\*** | 3.9 | **4.0\*** |

| wSD | *D* [$\times 10^{-3}$ mm²/s] | *f* [%] | *D\** [$\times 10^{-3}$ mm²/s] |
|---|---|---|---|
| **Least squares** | 0.10 | 6.2 | 24.9 |
| **Bayesian** | 0.09 | 4.9 | 5.1 |
| **IVIM-NET**optim | 0.06 | 2.4 | 15.8 |

| Mean treatment | *D* [$\times 10^{-3}$ mm²/s] | | *f* [%] | | *D\** [$\times 10^{-3}$ mm²/s] | |
|---|---|---|---|---|---|---|
| | Pre | Post | Pre | Post | Pre | Post |
| **Least squares** | **1.35\*\*** | **1.51\*\*** | 13.1 | 16.3 | 52 | 48 |
| **Bayesian** | 1.18 | 1.30 | 20.4 | 23.4 | 8.6 | 17.7 |
| **IVIM-NET**optim | **1.57\*\*** | **1.68\*\*** | **5.3\*\*** | **9.1\*\*** | 92 | 88 |

\* significantly (*P* < 0.05) better parameter map SNR compared to both of the other fitting approaches, determined by a two paired t-test, are printed bold.
\*\*significant (*P* < 0.05) changes between pre and post-treatment, determined by a paired t-test, are printed bold.

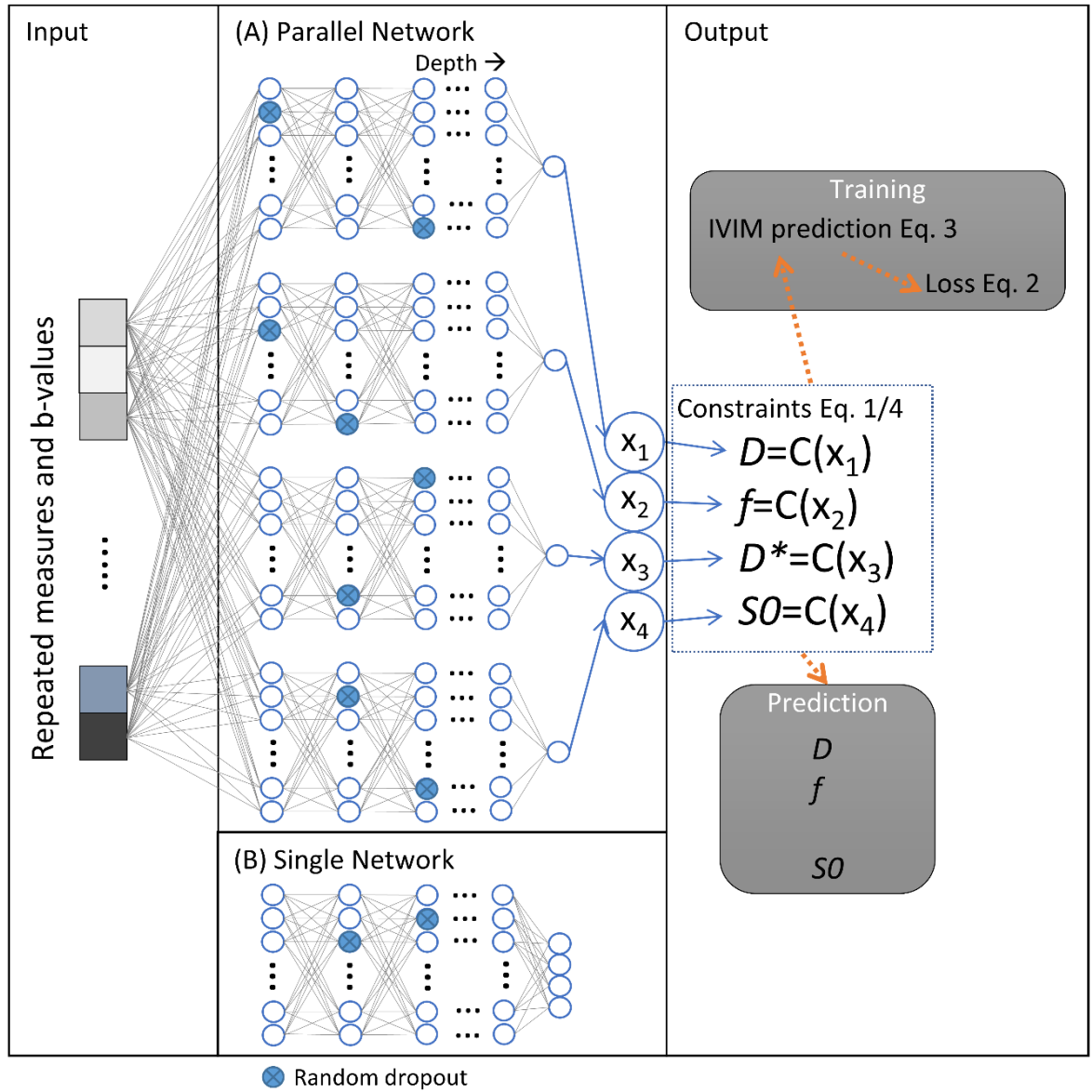

**Figure 1:** Representation of the PI-DNN with different hyperparameter options (Table 1). In this example, the input signal, consisting of the measured DWI signal, is fedforwarded either through (A) a parallel network design where each parameter is predicted by a separate fully connected set of hidden layers or (B) the original single fully-connected network design. The blue circles indicate an example of randomly selected neurons for dropout. In this example, the output layer consists of four neurons with either absolute (Eq. 1) or sigmoid activation functions (Eq. 4) whose values correspond to the IVIM parameters. Subsequently, the network predicts the IVIM signal (Eq. 3) which is used to compute the loss function (Eq. 2). With the loss function, the network trains the PI-DNN to give good estimates of the IVIM parameters.

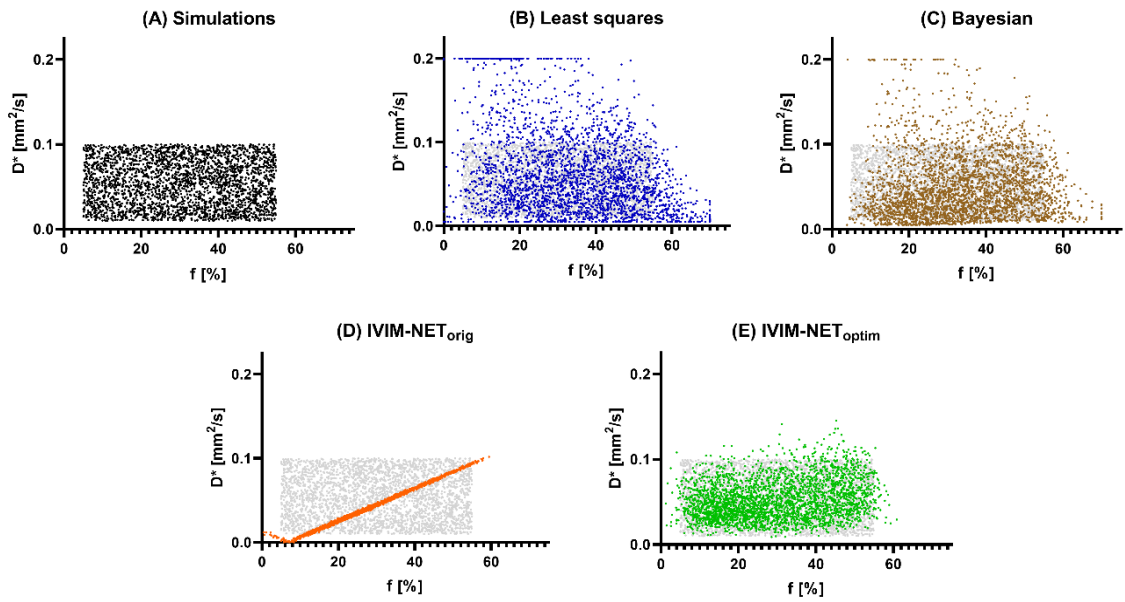

**Figure 2:** $D^*$ plotted against $f$ for (A) simulations, (B) Least squares, (C) Bayesian, (D) IVIM-NET$_{orig}$ and (E) IVIM-NET$_{optim}$. In all plots, the values of the simulations are presented in grey. The apparent patterns in the LS approach (many predictions at $D^* = 0.2$ mm$^2$/s) and IVIM-NET$_{orig}$ (the line flips at $D^* = 0$ mm$^2$/s) are a result of the fit constraints.

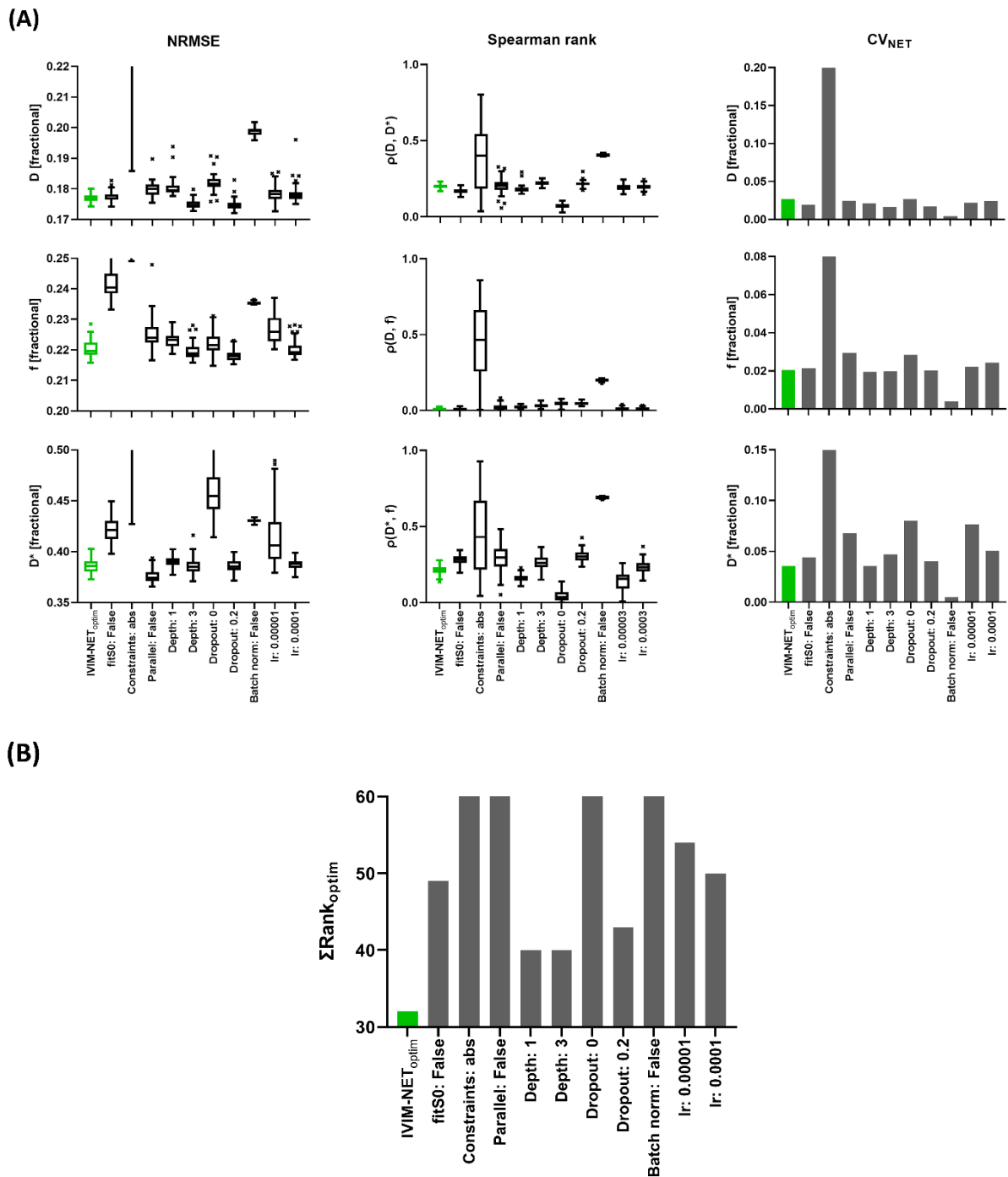

**Figure 3:** Normalized root-mean-square error (NRMSE; left), Spearman rank correlation coefficient ($\rho$; center) and normalized coefficient of variation ($CV_{NET}$; right) plots of the estimated IVIM parameters ($D$, $f$ and $D^*$) with a single parameter change for IVIM-NET$_{optim}$ (green) at SNR 20 for 50 repeated trainings. (B) The ranked plot of IVIM-NET$_{optim}$.

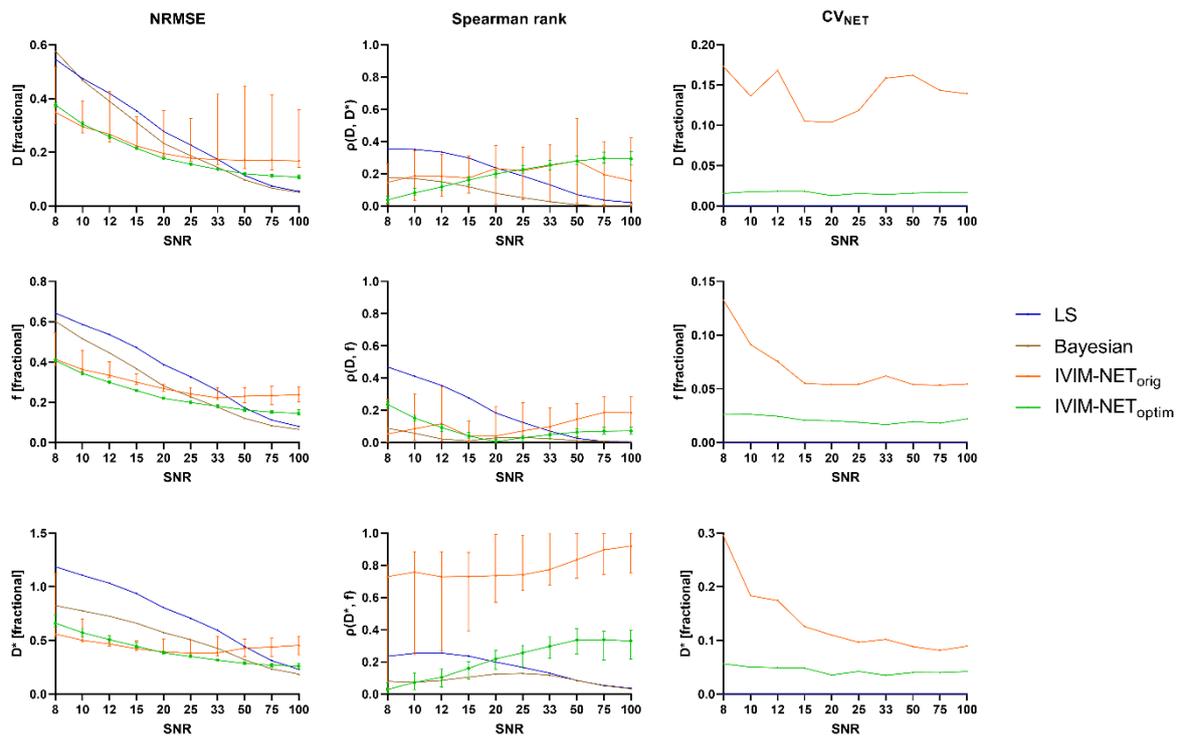

**Figure 4:** Normalized root-mean-square error (NRMSE; left), Spearman rank correlation coefficient ($\rho$; center) and normalized coefficient of variation ($CV_{NET}$; right) plots of the estimated IVIM parameters ($D$, $f$ and $D^*$) vs SNR for the LS approach (blue), Bayesian approach (brown), IVIM-NET$_{orig}$ (orange) and IVIM-NET$_{optim}$ (green) approaches to IVIM fitting. The 5 to 95 percentiles of IVIM-NET for 50 repeated trainings are plotted as error bars and show that IVIM-NET$_{orig}$ is highly inconsistent in producing IVIM parameters for multiple repeated trainings at all SNRs. IVIM-NET$_{optim}$ outperforms IVIM-NET$_{orig}$ for all SNRs. The LS and Bayesian approaches are superior at high SNRs.

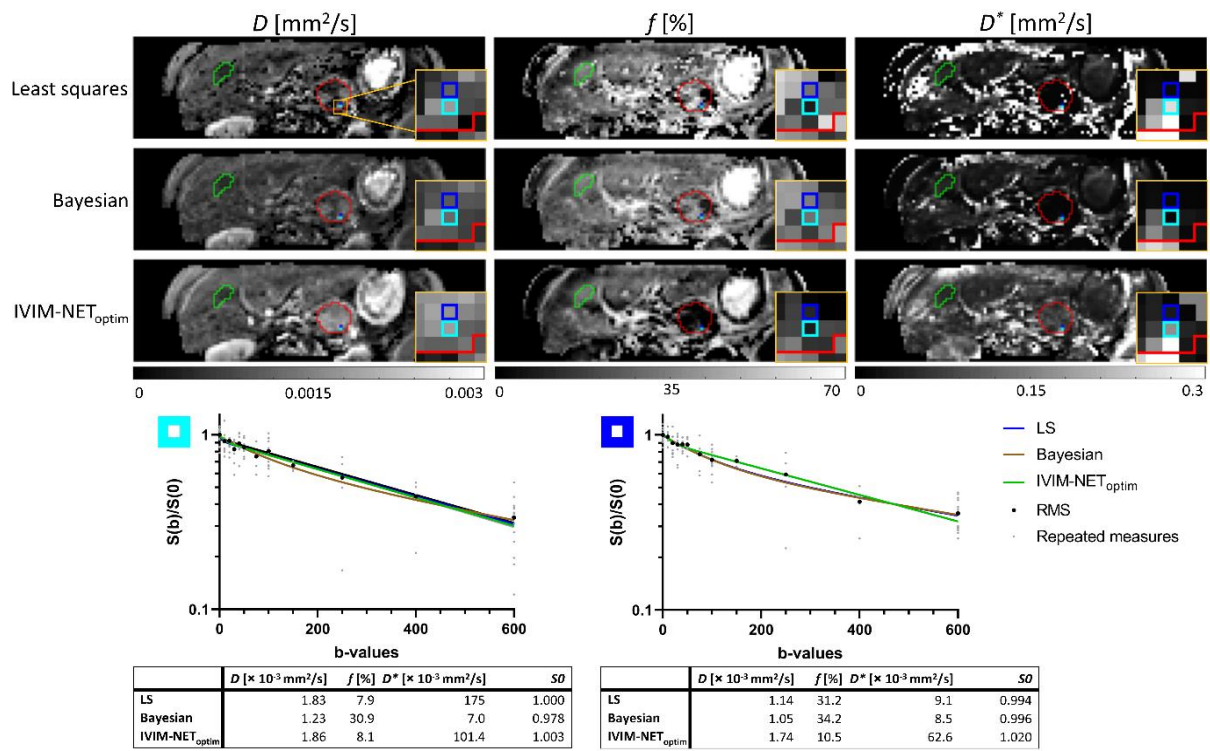

**Figure 5:** IVIM parameter maps (*D*, *f*, *D\**) of the LS approach, Bayesian approach and IVIM-NET$_{optim}$ of a PDAC patient of the test-retest cohort. The red ROI represents the PDAC, the two highlighted blue regions correlate to the voxels from the log-plots below. The yellow square zooms in on the two highlighted voxels in the tumor. In the plots, the small light grey dots are the repeated measures and the big black dots are the root-mean-squares of these repeated measures. The plot parameters are shown below. The light blue voxel (left plot) shows consistency in IVIM parameters for all three fitting approaches. Although the data is similar in the neighboring dark blue voxel (right plot) with a lower IVIM effect, the LS and Bayesian approaches compute a higher *f*, lower *D* and very low *D\** compared to their parameters in the light blue voxel. IVIM-NET$_{optim}$ shows more consistency in IVIM parameters between the two neighboring voxels with a lower *f*. In the parametric maps computed by IVIM-NET$_{optim}$, the tissues appear more homogeneous, whereas the LS approach shows noisier parameter maps, particularly around the tumor region.

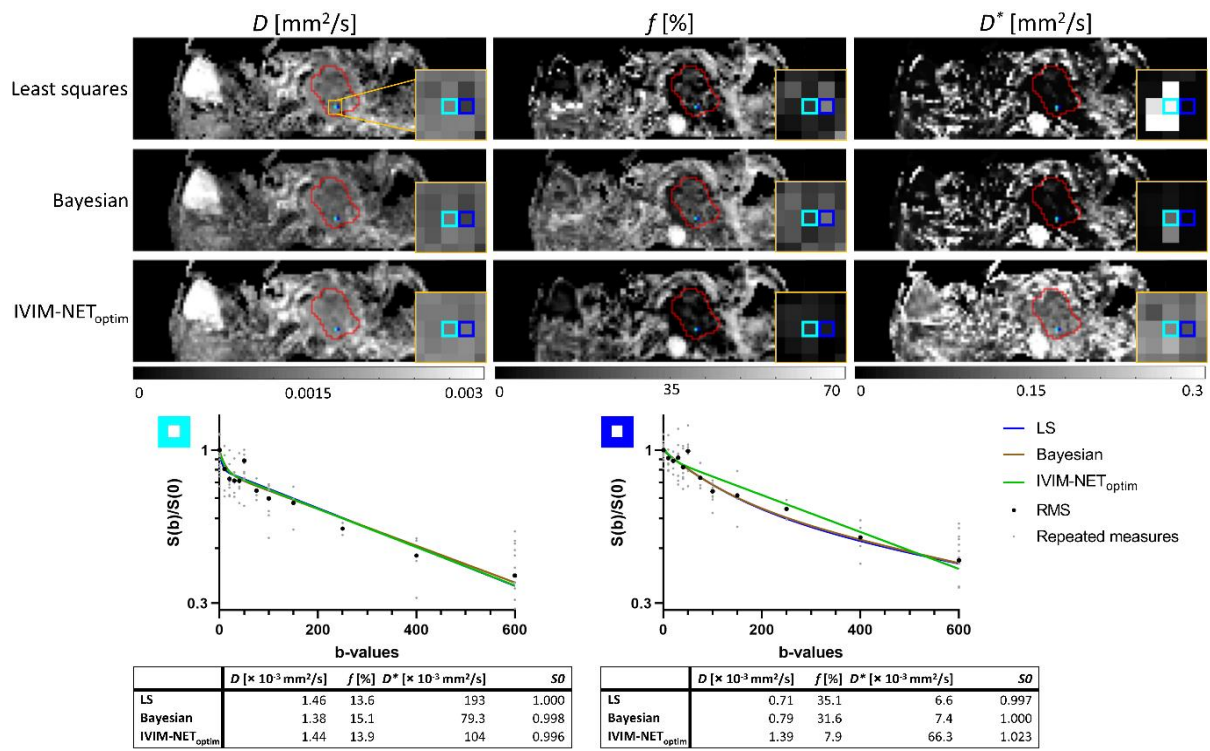

**Figure 6:** IVIM parameter maps (*D*, *f*, *D\**) of the LS approach, Bayesian approach and IVIM-NET$_{optim}$ of a PDAC patient of the treated cohort before CRT. The red ROI represents the PDAC and the green ROI represents the 2D homogenous liver tissue ROI. The two highlighted blue regions correlate to the voxels from the log-plots below. The yellow square zooms in on the two highlighted voxels in the tumor. In the plots, the small light grey dots are the repeated measures and the big black dots are the root-mean-squares of these repeated measures. The plot parameters are shown below. The light blue voxel (left plot) shows consistency in IVIM parameters between the LS approach and IVIM-NET$_{optim}$ with low *f*, and moderate *D* and *D\**, while the Bayesian approach shows higher *f*, lower *D* and very low *D\**. Although the data is similar in the neighboring dark blue voxel (right plot), the LS and Bayesian approaches compute a higher *f*, lower *D* and very low *D\** compared to their parameters in the light blue voxel. IVIM-NET$_{optim}$ shows more consistency in IVIM parameters between the two neighboring voxels. In the parametric maps computed by IVIM-NET$_{optim}$, the tissues appear more homogeneous, particularly in the liver, the kidneys and around the tumor ROI.

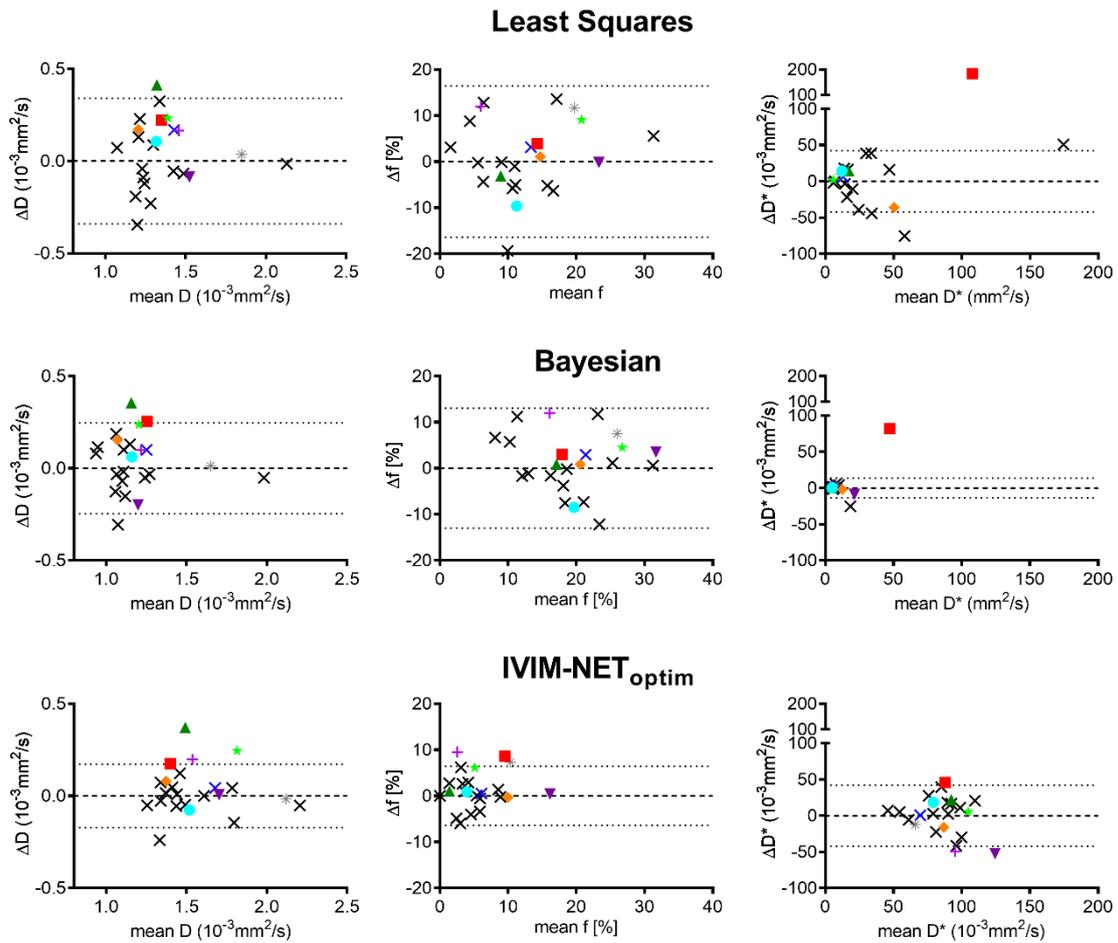

**Figure 7:** Bland-Altman plots of the LS, Bayesian and IVIM-NET$_{optim}$ approaches to IVIM fitting showing the mean and difference (Δ) between the intersession repeatability patients (black crosses) and the mean and Δ between pre and post-treatment patients (colored symbols) which represents the treatment effects. The dotted lines indicate the 95CI of the test-retest data. Colored measurements that exceed the 95CI were considered significant to treatment response.

## *Supporting Information 1: Simulations*

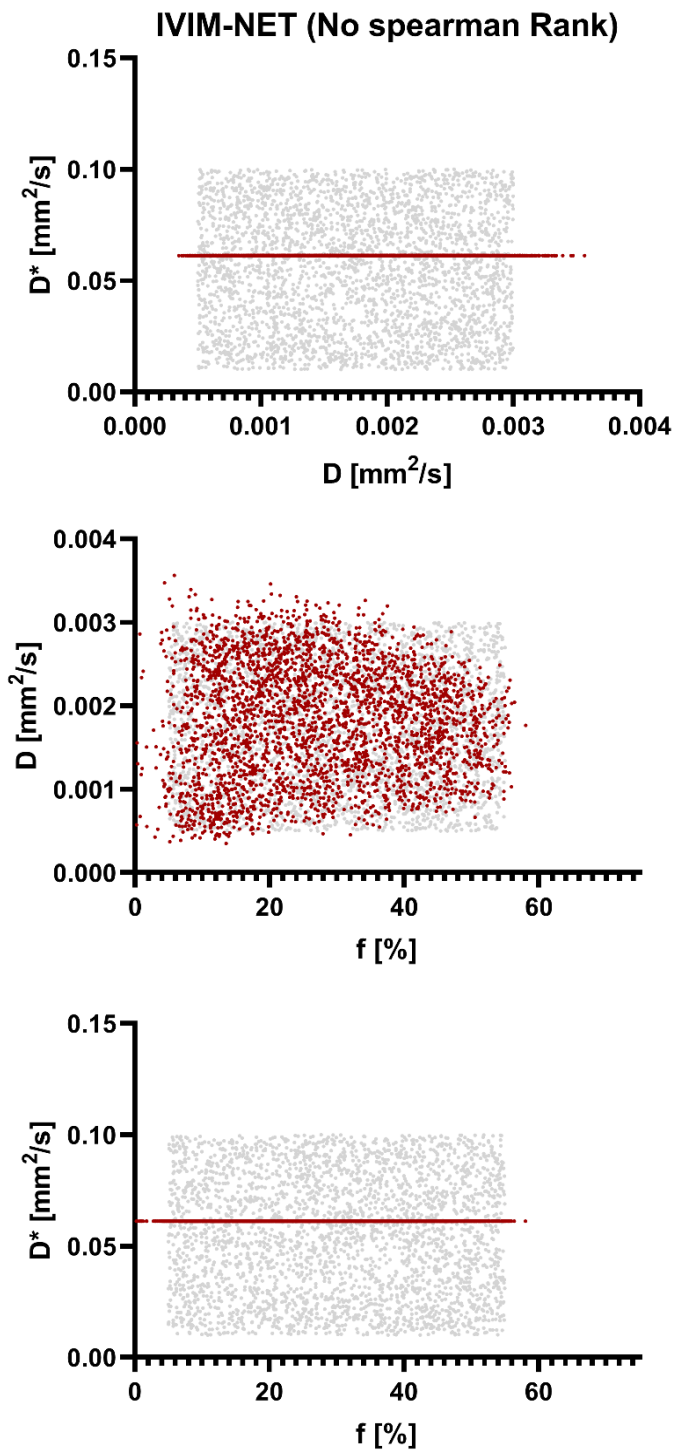

**Figure S1:** Plots of the estimated IVIM parameters ($D$, $f$, $D^*$) where no Spearman rank correlation coefficient ($\rho$) can be determined and is set to a $\rho$ of 1. In all plots, the values of the simulations are presented in grey.

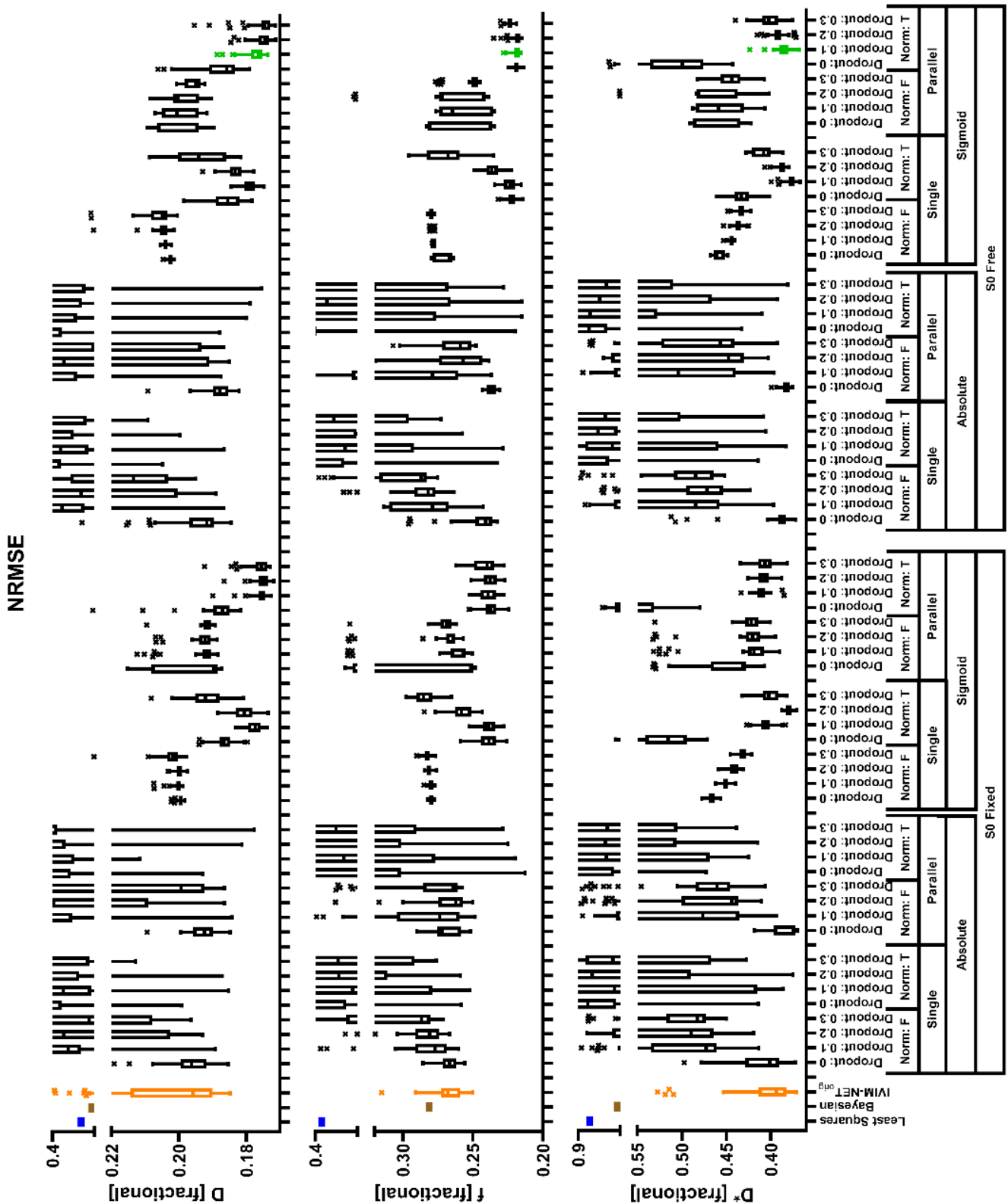

**Figure S2:** Normalized root-mean-square error (NRMSE) boxplots of the estimated IVIM parameters ($D$, $f$, $D^*$) that contain all hyperparameter combinations with a fixed learning rate set to $1 \times 10^{-4}$ and a fixed number of hidden layers set to 3 at SNR 20 for 50 repeated trainings. Highlighted in green is the intermediate step of IVIM-NET$_{optim}$. Left of each plot shows the LS approach (blue), Bayesian approach (brown) and IVIM-NET$_{orig}$ (orange; LR = $1 \times 10^{-3}$).

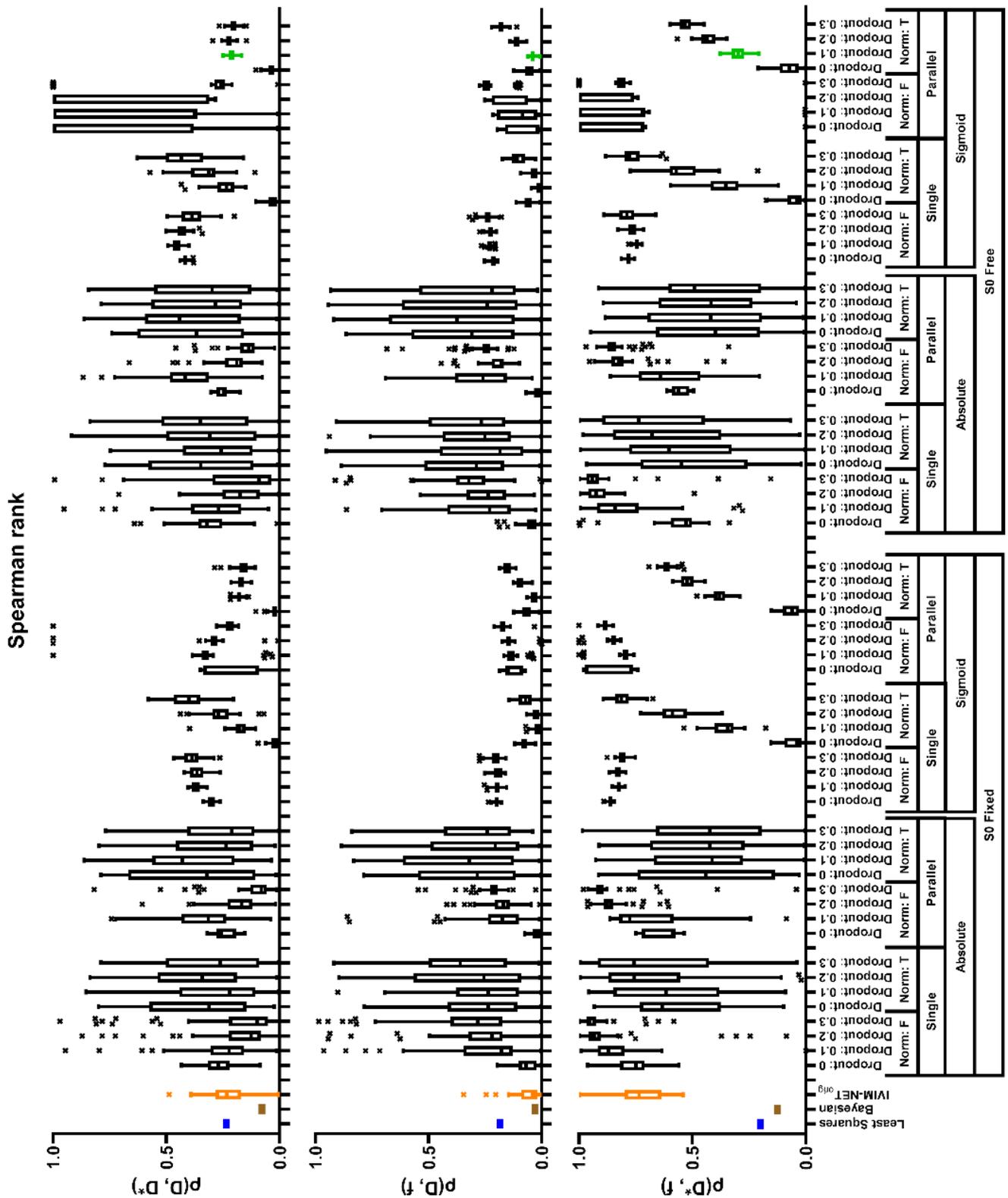

**Figure S3:** Spearman rank correlation coefficient ($\rho$) boxplots of the estimated IVIM parameters ($D$, $f$, $D^*$) that contain all hyperparameter combinations with a fixed learning rate set to $1 \times 10^{-4}$ and a fixed number of hidden layers set to 3 at SNR 20 for 50 repeated trainings. Highlighted in green is the intermediate step of IVIM-NET$_{optim}$. Left of each plot shows the LS approach (blue), Bayesian approach (brown) and IVIM-NET$_{orig}$ (orange; LR = $1 \times 10^{-3}$).

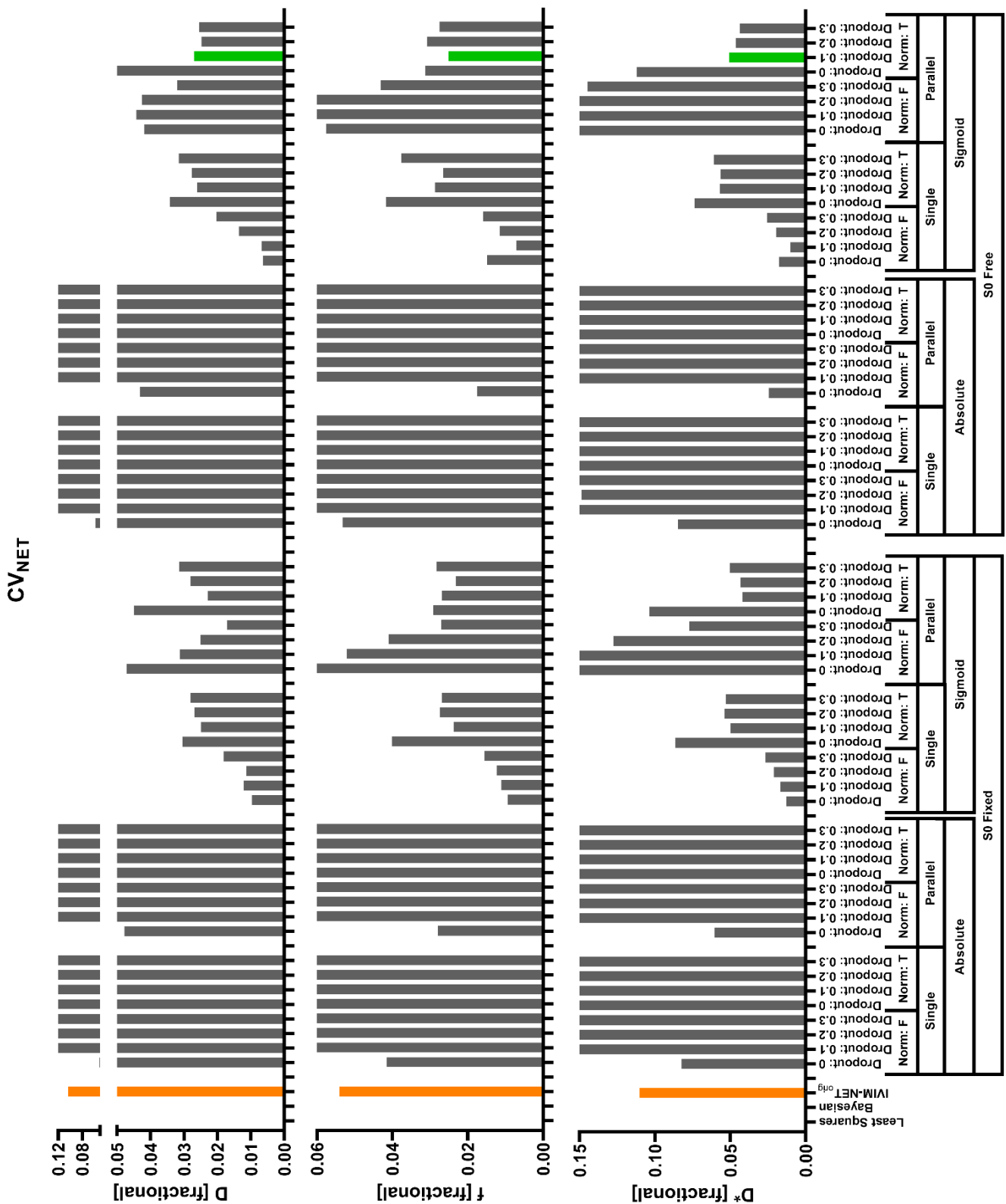

**Figure S4:** Normalized coefficient of variation (CV$_{NET}$) plots of the estimated IVIM parameters ($D$, $f$, $D*$) that contain all hyperparameter combinations with a fixed learning rate set to $1 \times 10^{-4}$ and a fixed number of hidden layers set to 3 at SNR 20 for 50 repeated trainings. Highlighted in green is the intermediate step of IVIM-NET$_{optim}$. Left of each plot shows the LS approach (blue), Bayesian approach (brown) and IVIM-NET$_{orig}$ (orange; LR = $1 \times 10^{-3}$).

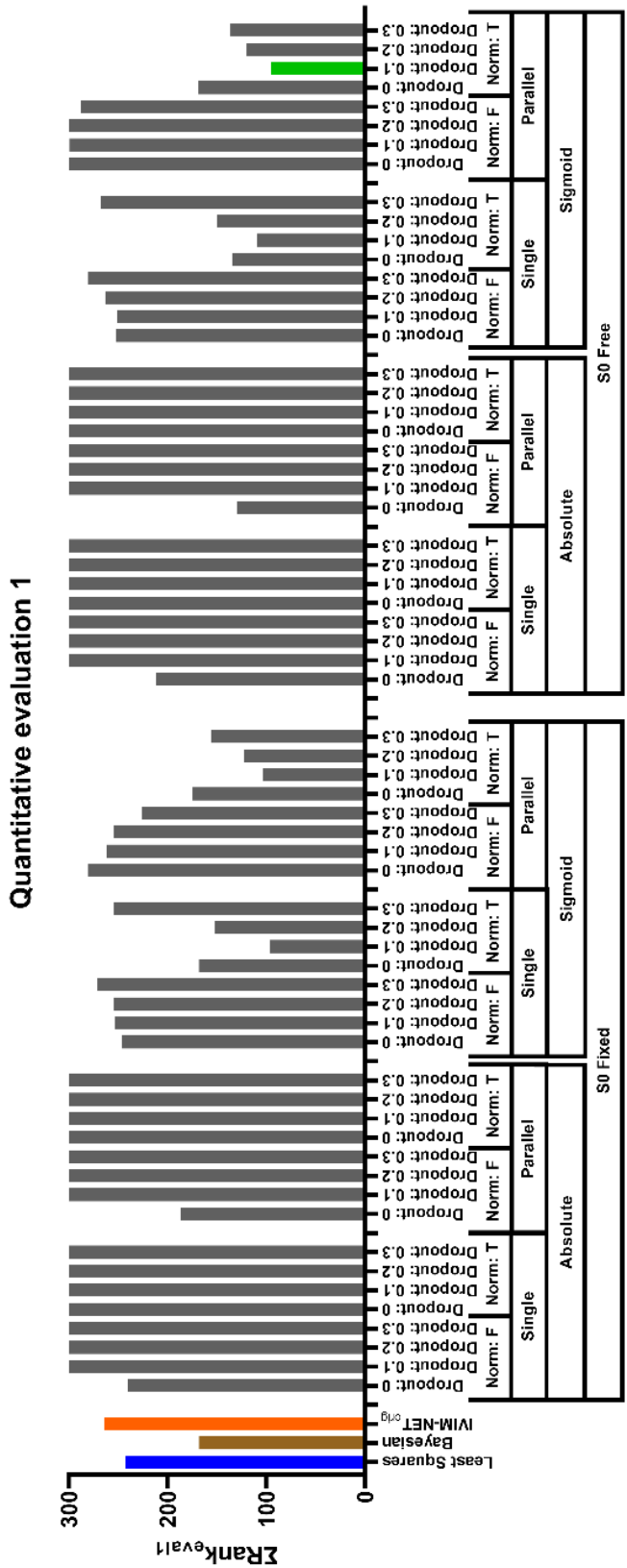

**Figure S5:** Ranked plots of the metrics (NRMSE, $\rho$ and $CV_{NET}$) of evaluation 1 that contain all hyperparameter combinations with a fixed learning rate set to $1 \times 10^{-4}$ and a fixed number of hidden layers set to 3 at SNR 20 for 50 repeated trainings. Highlighted in green is the intermediate step of IVIM-NET$_{optim}$. Left of each plot shows the LS approach (blue), Bayesian approach (brown) and IVIM-NET$_{orig}$ (orange; LR = $1 \times 10^{-3}$).

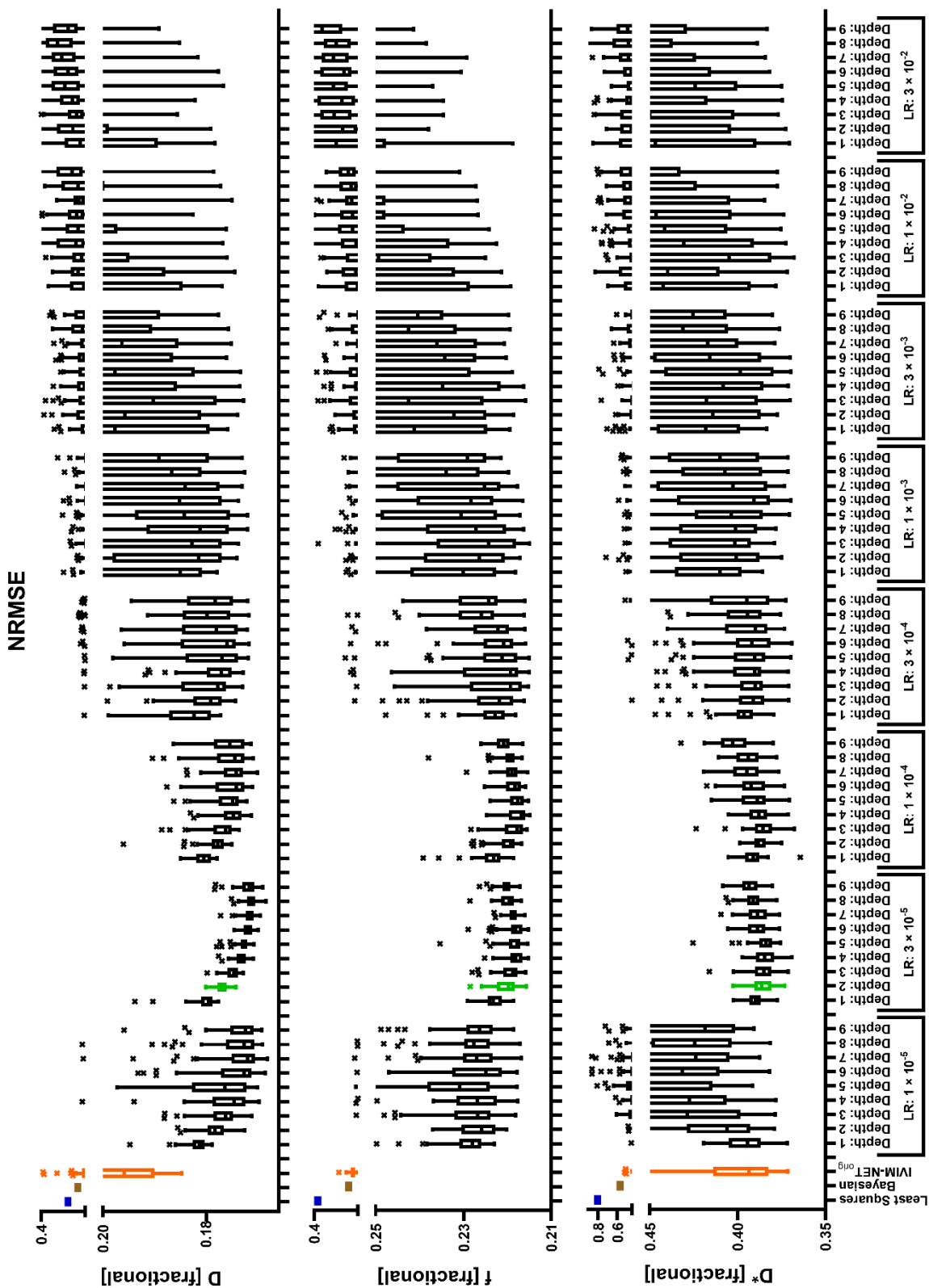

**Figure S6:** Normalized root-mean-square error (NRMSE) boxplots of the estimated IVIM parameters ($D$, $f$, $D^*$) of the second evaluation for different LR and number of hidden layers, with fixed hyperparameters of extra fitting parameter $S0$, sigmoid activation functions, a parallel network architecture, 10% dropout and batch normalization at SNR 20 for 50 repeated trainings. Highlighted in green is IVIM-NET$_{optim}$. Left of each plot shows the LS approach (blue) and Bayesian approach (brown) and IVIM-NET$_{orig}$ (orange).

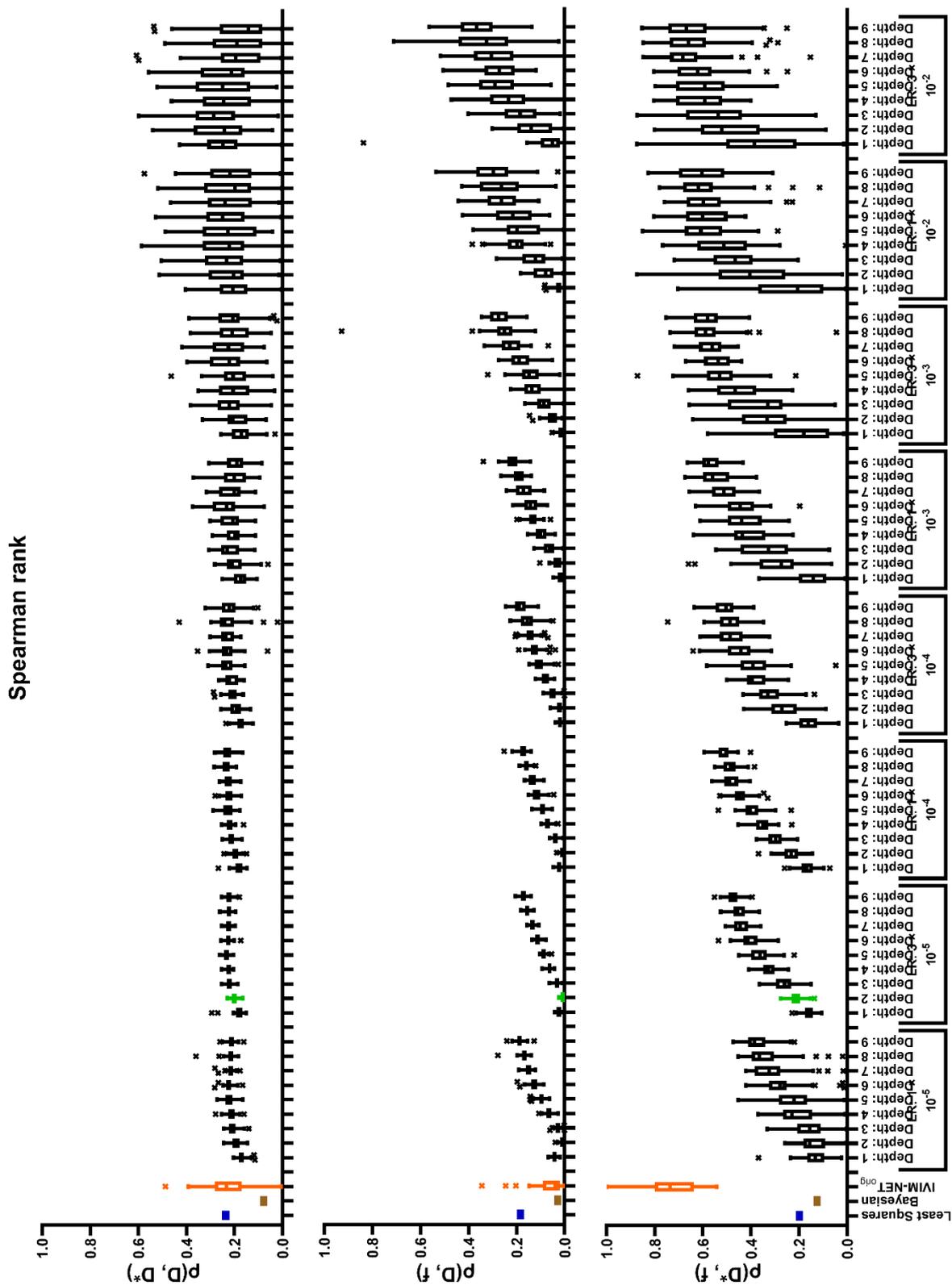

**Figure S7:** Spearman rank correlation coefficient ($\rho$) boxplots of the estimated IVIM parameters ($D$, $f$, $D^*$) of the second evaluation for different LR and number of hidden layers, with fixed hyperparameters of extra fitting parameter $S0$, sigmoid activation functions, a parallel network architecture, 10% dropout and batch normalization at SNR 20 for 50 repeated trainings. Highlighted in green is IVIM-NET$_{optim}$. Left of each plot shows the LS approach (blue) and Bayesian approach (brown) and IVIM-NET$_{orig}$ (orange).

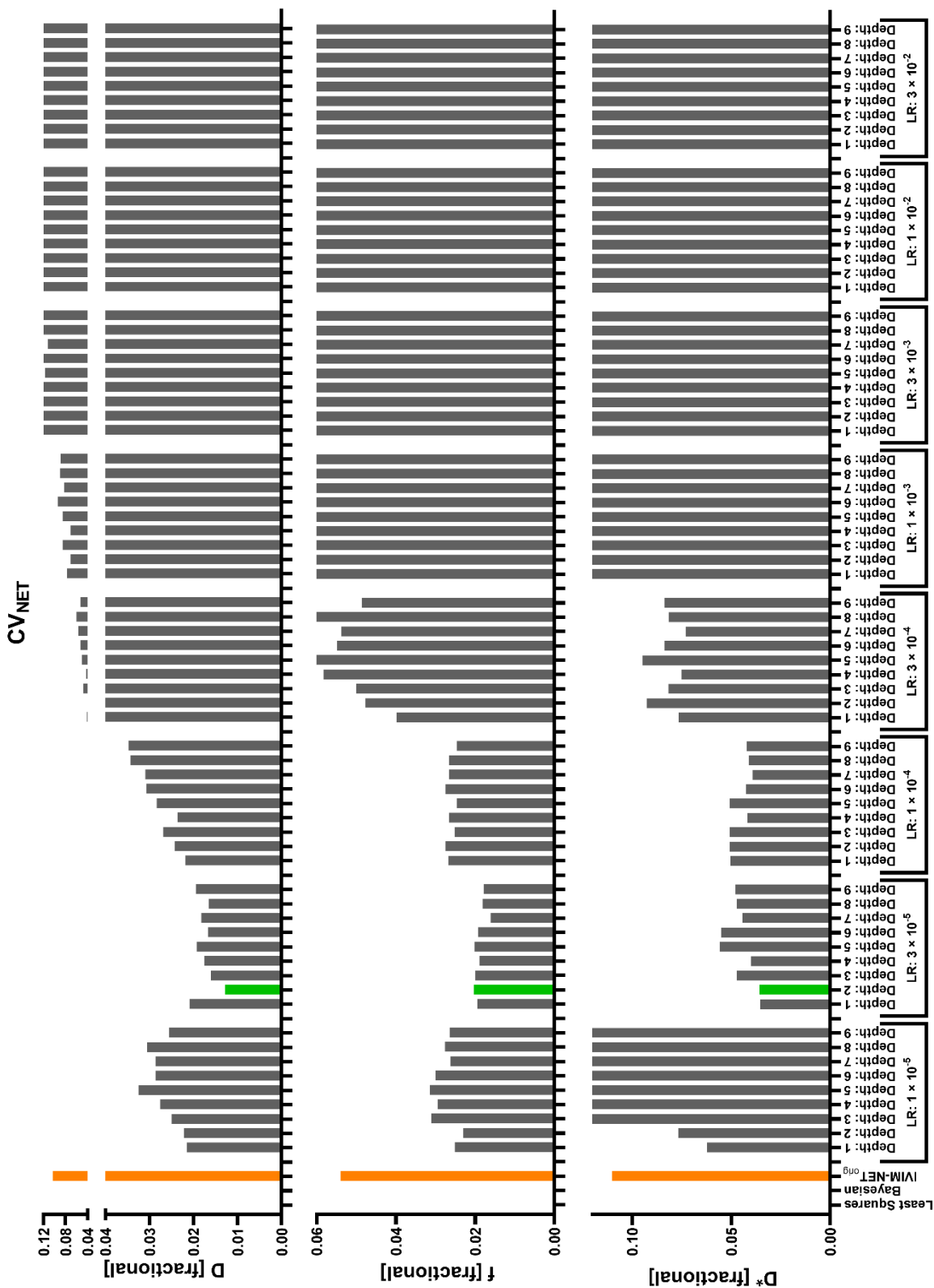

**Figure S8:** Normalized coefficient of variation (CV$_{NET}$) plots of the estimated IVIM parameters ($D$, $f$, $D^*$) of the second evaluation for different LR and number of hidden layers, with fixed hyperparameters of extra fitting parameter $S0$, sigmoid activation functions, a parallel network architecture, 10% dropout and batch normalization at SNR 20 for 50 repeated trainings. Highlighted in green is IVIM-NET$_{optim}$. Left of each plot shows the LS approach (blue) and Bayesian approach (brown) and IVIM-NET$_{orig}$ (orange).

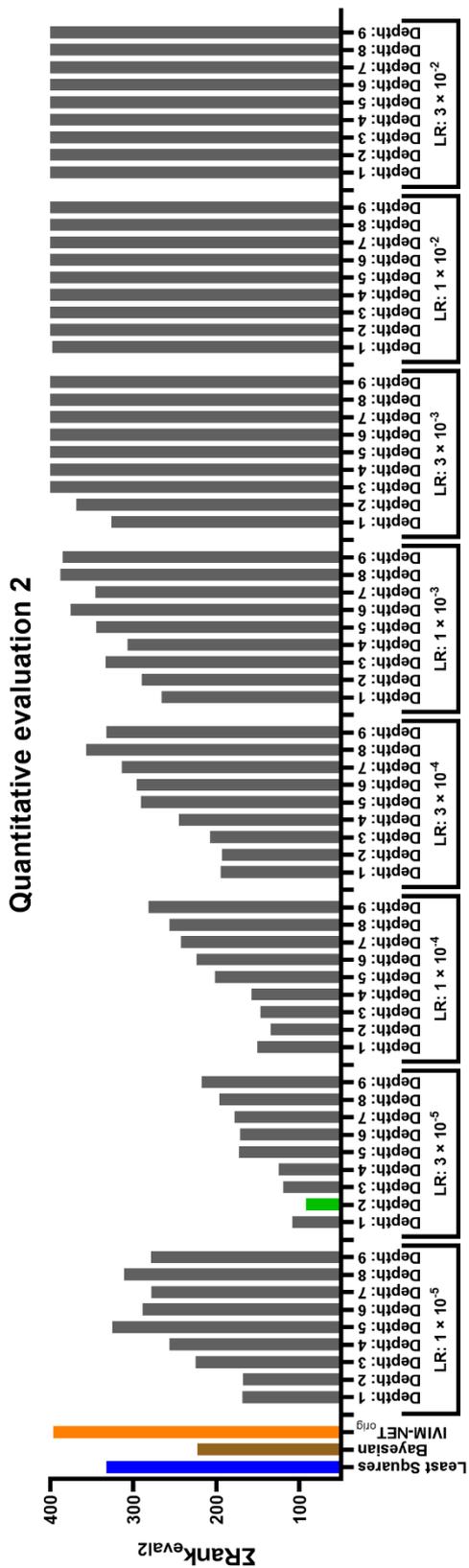

**Figure S9:** Ranked plots of the metrics (NRMSE, $\rho$ and $CV_{NET}$) of evaluation 2 for different LR and number of hidden layers, with fixed hyperparameters of extra fitting parameter $S0$, sigmoid activation functions, a parallel network architecture, 10% dropout and batch normalization at SNR 20 for 50 repeated trainings. Highlighted in green is IVIM-NET$_{optim}$. Left of each plot shows the LS approach (blue) and Bayesian approach (brown) and IVIM-NET$_{orig}$ (orange).

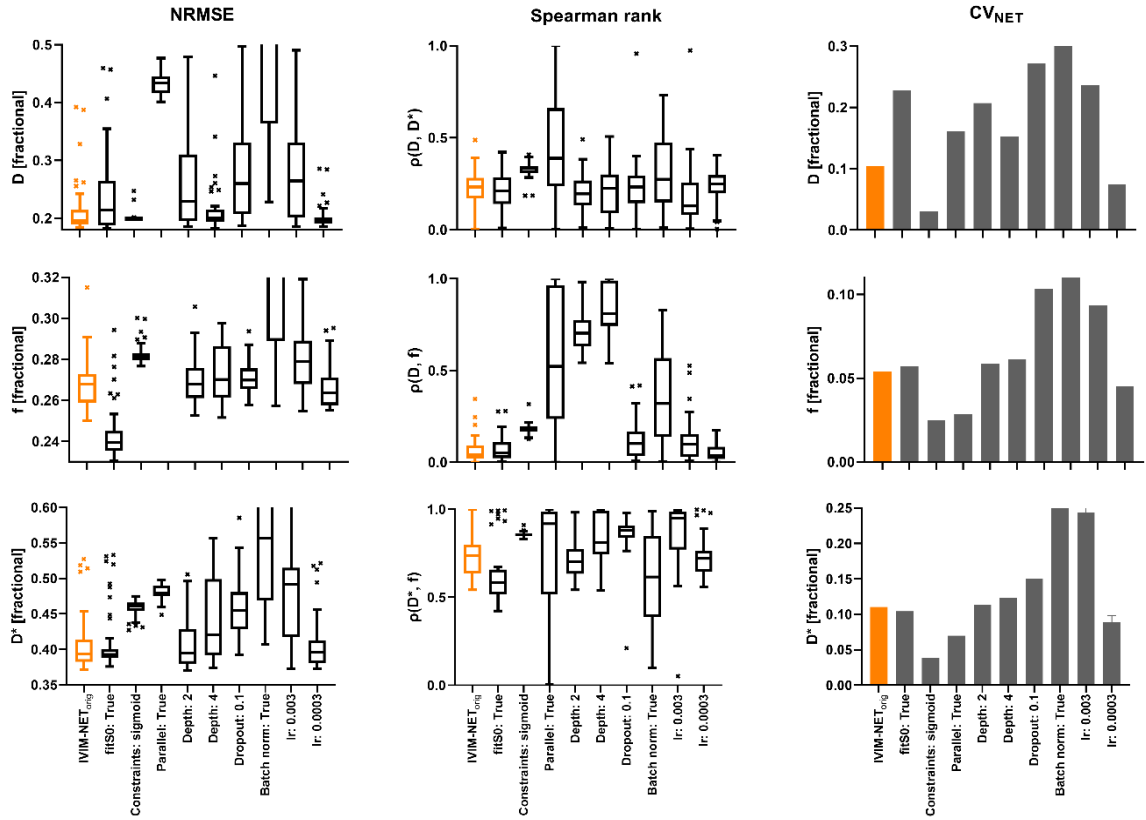

**Figure S10:** Normalised root-mean-square error (NRMSE; left), Spearman rank correlation coefficient ($\rho$; center) and normalized coefficient of variation ($CV_{NET}$; right) plots of the estimated IVIM parameters ($D$, $f$ and $D^*$) with a single parameter change for IVIM-NET$_{orig}$ (orange) at SNR 20 for 50 repeated trainings. The $\rho(D^*,f)$ remains substantial for single deviations from IVIM-NET$_{orig}$.

### Supporting Information 2: Verification in patients with PDAC

For simplicity, we have made a table overview (Table S1) of every parameter map of the Supporting Information. As in Figures 5 and 6 of our manuscript, these Supporting Information Figures S11-S20 show 'IVIM parameter maps ($D$, $f$, $D^*$) of the LS approach, Bayesian approach and IVIM-NET$_{optim}$ of a PDAC patient of the treated cohort before CRT and the test-retest cohort. The red ROI represents the PDAC and the green ROI represents homogenous 2D liver tissue ROI. The two highlighted blue regions correlate to the voxels from the plots below. The yellow square zooms in on the two highlighted voxels. In the plots, the small light grey dots are the repeated measures and the big black dots are the root-mean-squares of these repeated measures. The plot parameters are shown below.'

**Table S1:** Overview of the parameter maps of Supporting Information Figures S11-S20.

| Figure | Remark |
|---|---|
| S11 | Highlighted voxels in the liver. The light blue voxel (left plot) shows consistency in IVIM parameters between the LS approach and IVIM-NET$_{optim}$ with low $f$, and moderate $D$ and $D^*$, while the Bayesian approach shows higher $f$, lower $D$ and very low $D^*$. The neighboring dark blue voxel (right plot) shows no diffusion ($D = 0$ mm$^2$/s), a very high $f$ and very low $D^*$ for the LS and Bayesian approaches. IVIM-NET$_{optim}$ shows more consistency in IVIM parameters between the two neighboring voxels. The LS and Bayesian approaches show noisier parameter maps, particularly in the liver and around the tumor region. |
| S12 | Highlighted voxels in the tumor. The light blue voxel (left plot) shows indifferent IVIM parameters for all three fitting approaches. Although the data is similar in the neighboring dark blue voxel (right plot), the LS and Bayesian approaches compute a higher $f$, lower $D$ (with $D = 0$ mm$^2$/s for the LS approach) and very low $D^*$ (to the lower bound of $D^* = 5.0 \times 10^{-3}$ mm$^2$/s for the LS approach) compared to their parameters in the light blue voxel. IVIM-NET$_{optim}$ shows more consistency in IVIM parameters between the two neighboring voxels. The LS and Bayesian approaches show noisier parameter maps, particularly in the liver and around the tumor region. Note that there is an artifact, which can be seen best in the middle part of the liver. |
| S13 | Highlighted voxels in the liver. The light blue voxel (left plot) shows consistency in IVIM parameters for all three fitting approaches with a high IVIM effect. Although the data is similar in the neighboring dark blue voxel (right plot), the LS and Bayesian approaches compute a higher $f$, and lower $D$ and $D^*$ compared to their parameters in the light blue voxel. IVIM-NET$_{optim}$ shows more consistency in IVIM parameters between the two neighboring voxels. The LS and Bayesian approaches show noisier parameter maps, particularly in the liver and around the tumor region. |
| S14 | Highlighted voxels in the tumor. The light blue voxel (left plot) shows consistency in IVIM parameters for all three fitting approaches. Although the data is similar in the neighboring dark blue voxel (right plot), the LS and Bayesian approaches compute a higher $f$, and lower $D$ and very low $D^*$ (to the lower bound of $D^* = 5.0 \times 10^{-3}$ mm$^2$/s for the LS approach) compared to their parameters in the light blue voxel. IVIM-NET$_{optim}$ shows more consistency in IVIM parameters between the two neighboring voxels. The LS and Bayesian approaches show noisier parameter maps, particularly in the liver, kidneys and around the tumor region. |
| S15 | Highlighted voxels in the tumor. The light blue voxel (left plot) shows consistency in IVIM parameters between the LS approach and IVIM-NET$_{optim}$ with low $f$, and moderate $D$ and $D^*$, while the Bayesian approach shows higher $f$, lower $D$ and very low $D^*$. Although the data is similar in the neighboring dark blue voxel (right plot), the LS and Bayesian approaches compute a higher $f$, and lower $D$ and very low $D^*$ (to the lower bound of $D^* = 5.0 \times 10^{-3}$ mm$^2$/s) compared to their parameters in the light blue voxel. IVIM-NET$_{optim}$ shows more consistency in IVIM parameters between the two neighboring voxels. The LS and Bayesian approaches show noisier parameter maps, particularly in the kidneys and around the tumor region. |
| S16 | Highlighted voxels in the tumor. The light blue voxel (left plot) shows consistency in IVIM parameters for all three fitting approaches. Although the data is similar in the neighboring dark blue voxel (right plot) with a lower IVIM effect, the LS and Bayesian approaches compute a higher $f$, and lower $D$ and very low $D^*$ (to the lower bound of $D^* = 5.0 \times 10^{-3}$ mm$^2$/s) compared to their parameters in the light blue voxel with a lower $f$. IVIM-NET$_{optim}$ shows more consistency in IVIM parameters between the two neighboring voxels with a lower $f$. The LS and Bayesian approaches show noisier parameter maps, particularly around the tumor region. Note that the LS approach has a very high $D^*$ (to the upper bound of $D^* = 200 \times 10^{-3}$ mm$^2$/s) in the blue voxel, while in the neighboring dark blue voxel it has a very low $D^*$ (to the lower bound of $D^* = 5.0 \times 10^{-3}$ mm$^2$/s). |
| S17 | Highlighted voxels in the liver. The light blue voxel (left plot) shows consistency in IVIM parameters for all three fitting approaches with a high IVIM effect. Although the data is similar in the neighboring dark blue voxel (right plot) with a lower IVIM effect, the LS and Bayesian approaches compute a higher $f$, and lower $D$ and $D^*$ compared to their parameters in the light blue voxel. IVIM-NET$_{optim}$ shows more consistency in IVIM parameters between the two neighboring voxels with a lower $f$. The LS and Bayesian approaches show noisier parameter maps, particularly in the liver. |
| S18 | Highlighted voxels in the kidneys. The light blue voxel (left plot) shows consistency in IVIM parameters for all three fitting approaches. Although the data is similar in the neighboring dark blue voxel (right plot) with a lower IVIM effect, the Bayesian approaches compute a higher $f$, and lower $D$ and very low $D^*$ compared to its parameters in the light blue voxel. IVIM-NET$_{optim}$ and the LS approach show more consistency in IVIM parameters between the two neighboring voxels with a lower $f$. The LS and Bayesian approaches show noisier parameter maps, particularly in the liver, kidneys and around the tumor region. |
| S19 | Highlighted voxels in the liver. The light blue voxel (left plot) shows consistency in IVIM parameters between the LS approach and IVIM-NET$_{optim}$ with low $f$, lower $D$ and moderate $D$ and $D^*$, while the Bayesian approach shows higher $f$, lower $D$ and very low $D^*$. Although the data is similar in the neighboring dark blue voxel (right plot), the LS and Bayesian approaches compute a higher $f$, and lower $D$ and very low $D^*$ compared to their parameters in the light blue voxel. IVIM-NET$_{optim}$ shows more consistency in IVIM parameters between the two neighboring voxels. The LS and Bayesian approaches show noisier parameter maps, particularly in the liver. |
| S20 | Highlighted voxels in the liver. The light blue voxel (left plot) shows consistency in IVIM parameters for all three fitting approaches. Although the data is similar in the neighboring dark blue voxel (right plot) with a lower IVIM effect, the Bayesian approaches compute a higher $f$, and lower $D$ and very low $D^*$ compared to its parameters in the light blue voxel. IVIM-NET$_{optim}$ and the LS approach show more consistency in IVIM parameters between the two neighboring voxels with a lower $f$. The LS approach shows noisier parameter maps, particularly in the liver, kidneys and around the tumor region. Note that although the LS approach has a very high D* (to the upper bound of $D^* = 200 \times 10^{-3}$ mm$^2$/s) in both voxels, the LS and IVIM-NET$_{optim}$ show the same plots. |

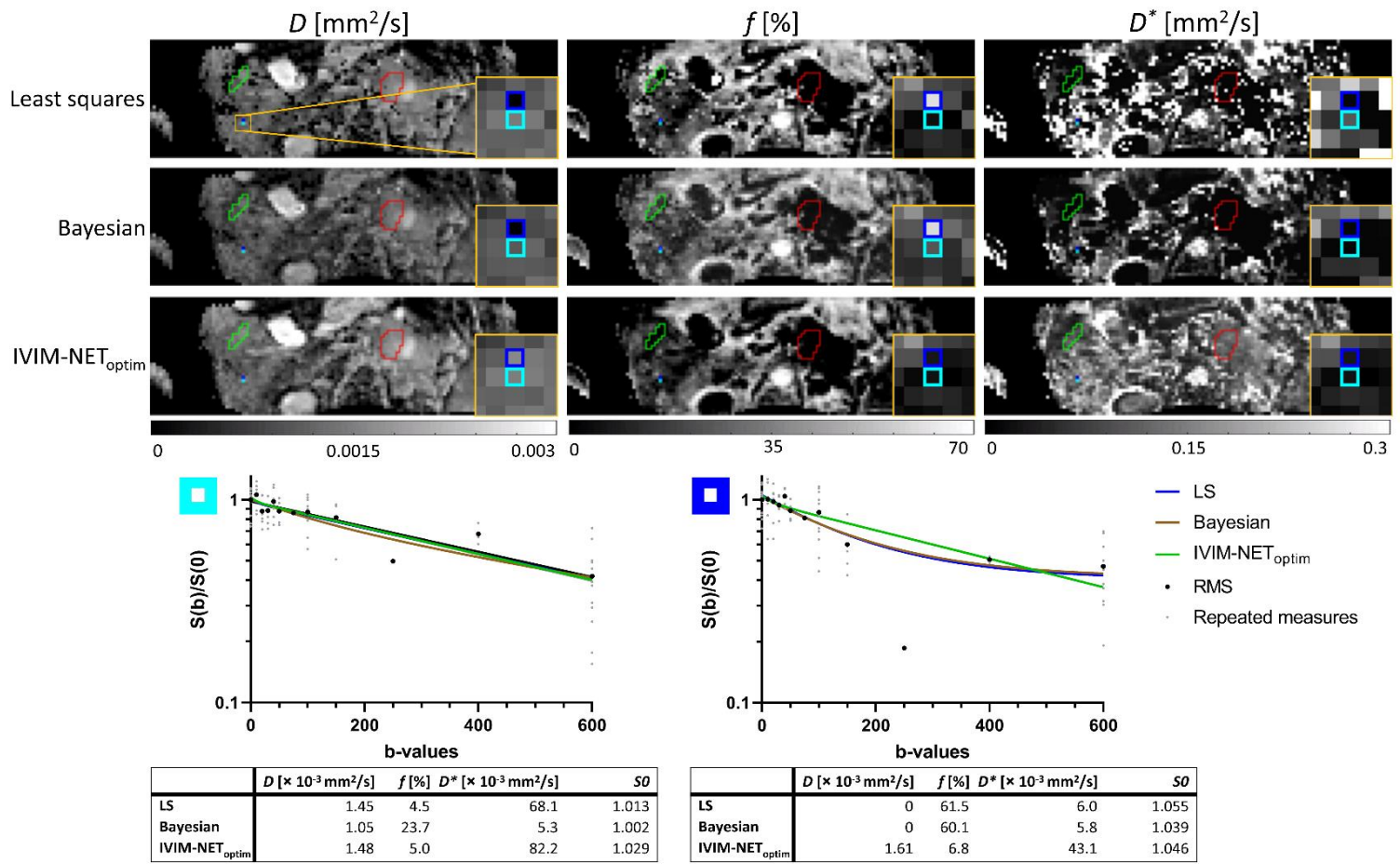

**Figure S11:** See Table S1.

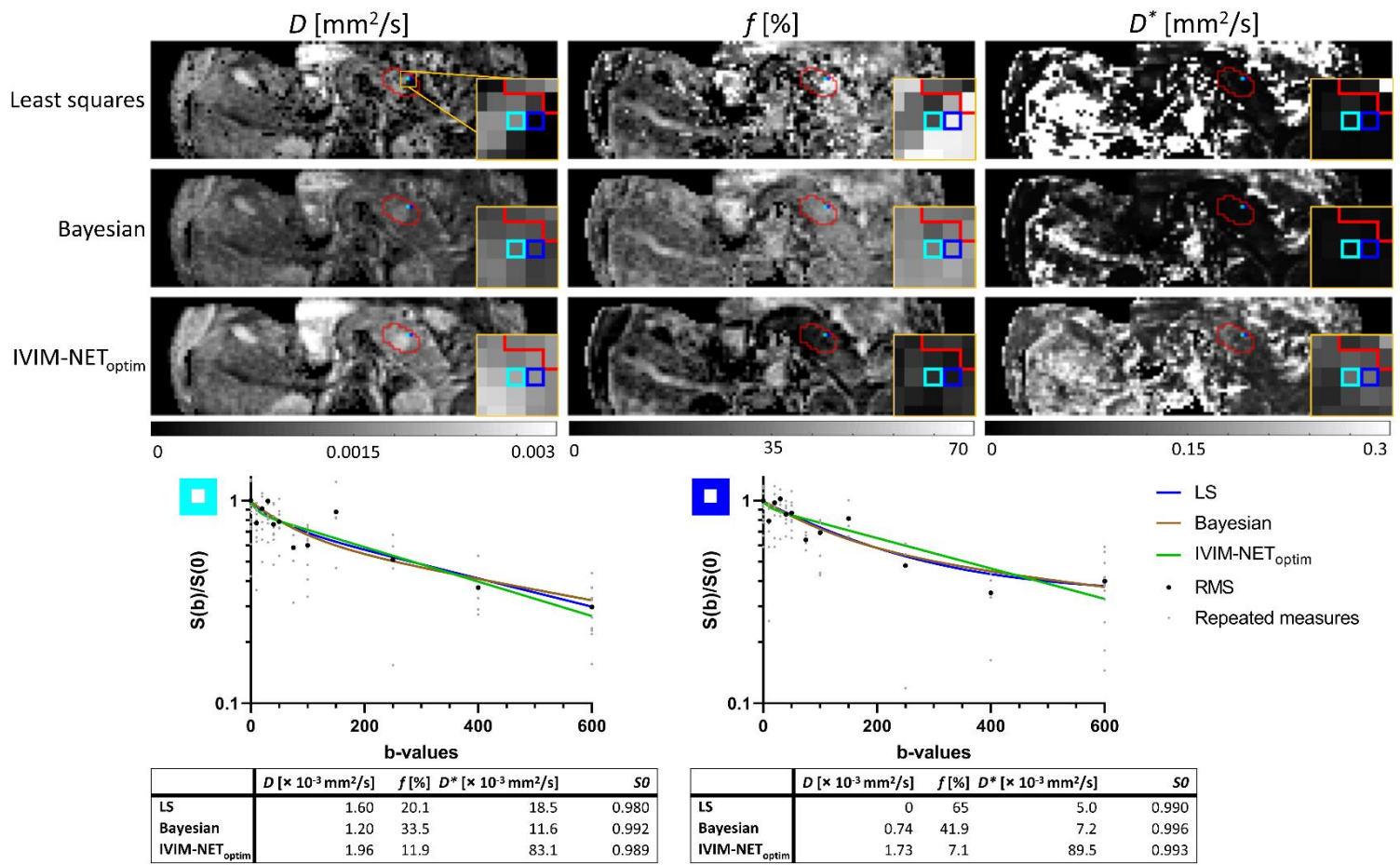

**Figure S12:** See Table S1.

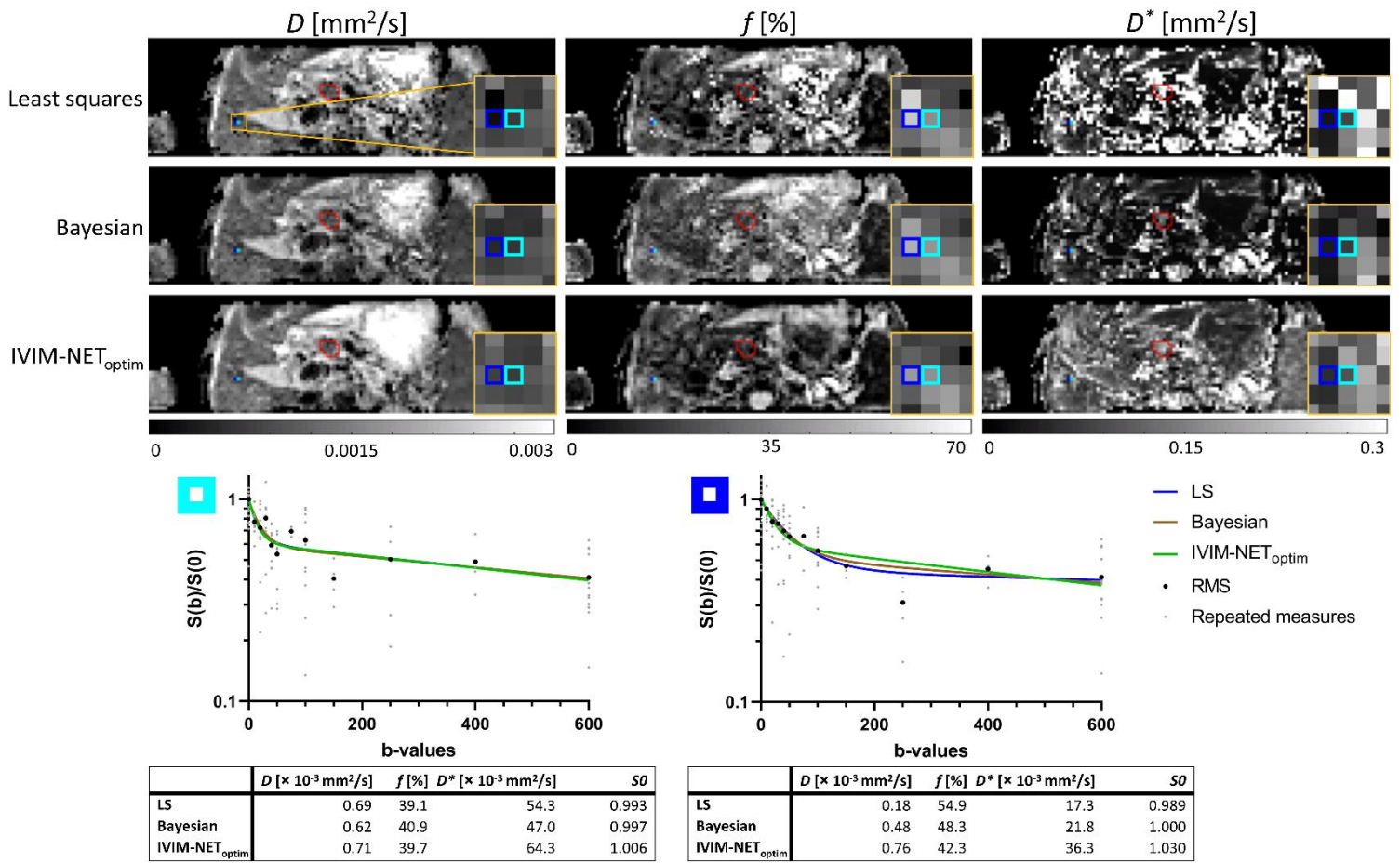

**Figure S13:** See Table S1.

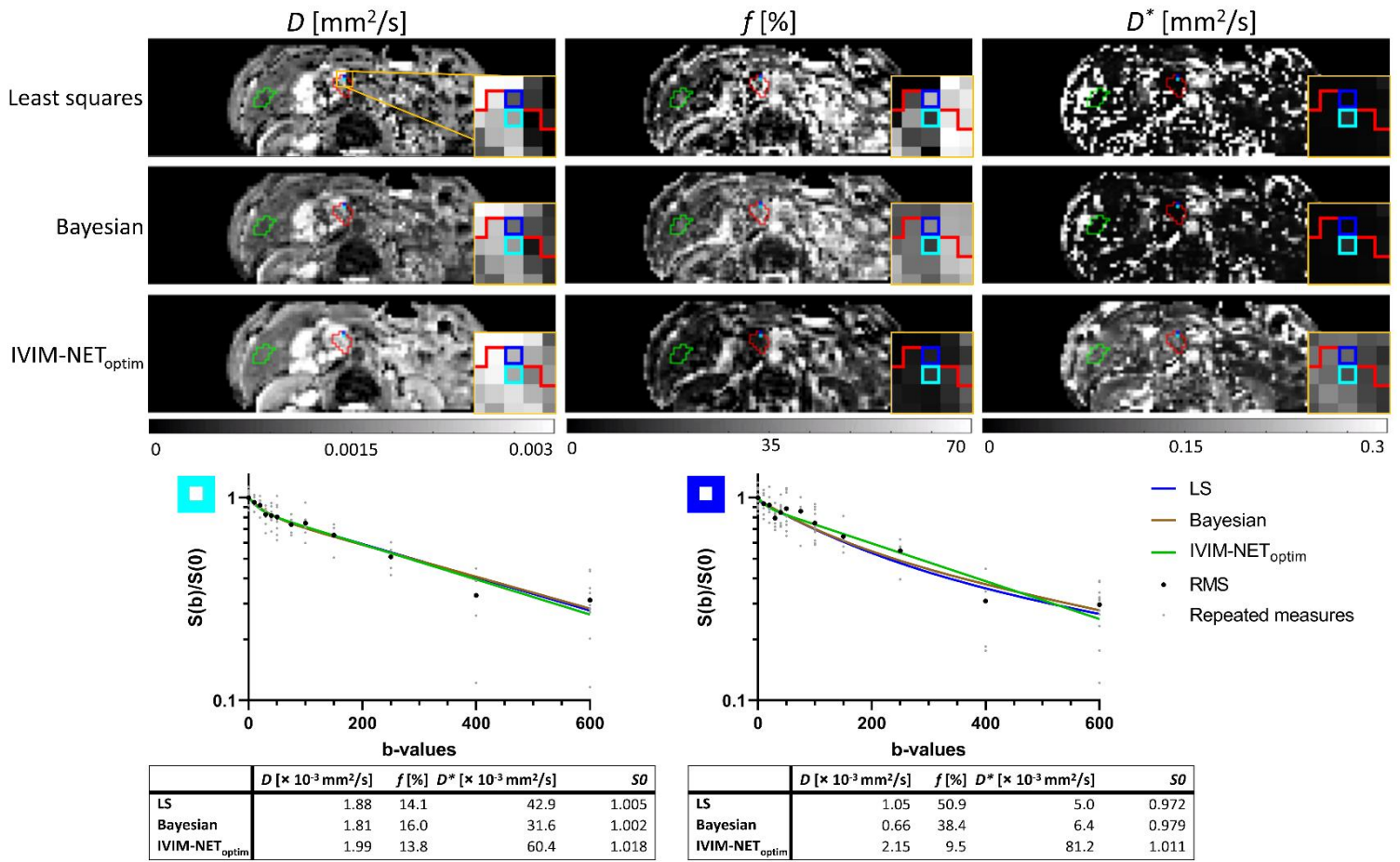

**Figure S14:** See Table S1.

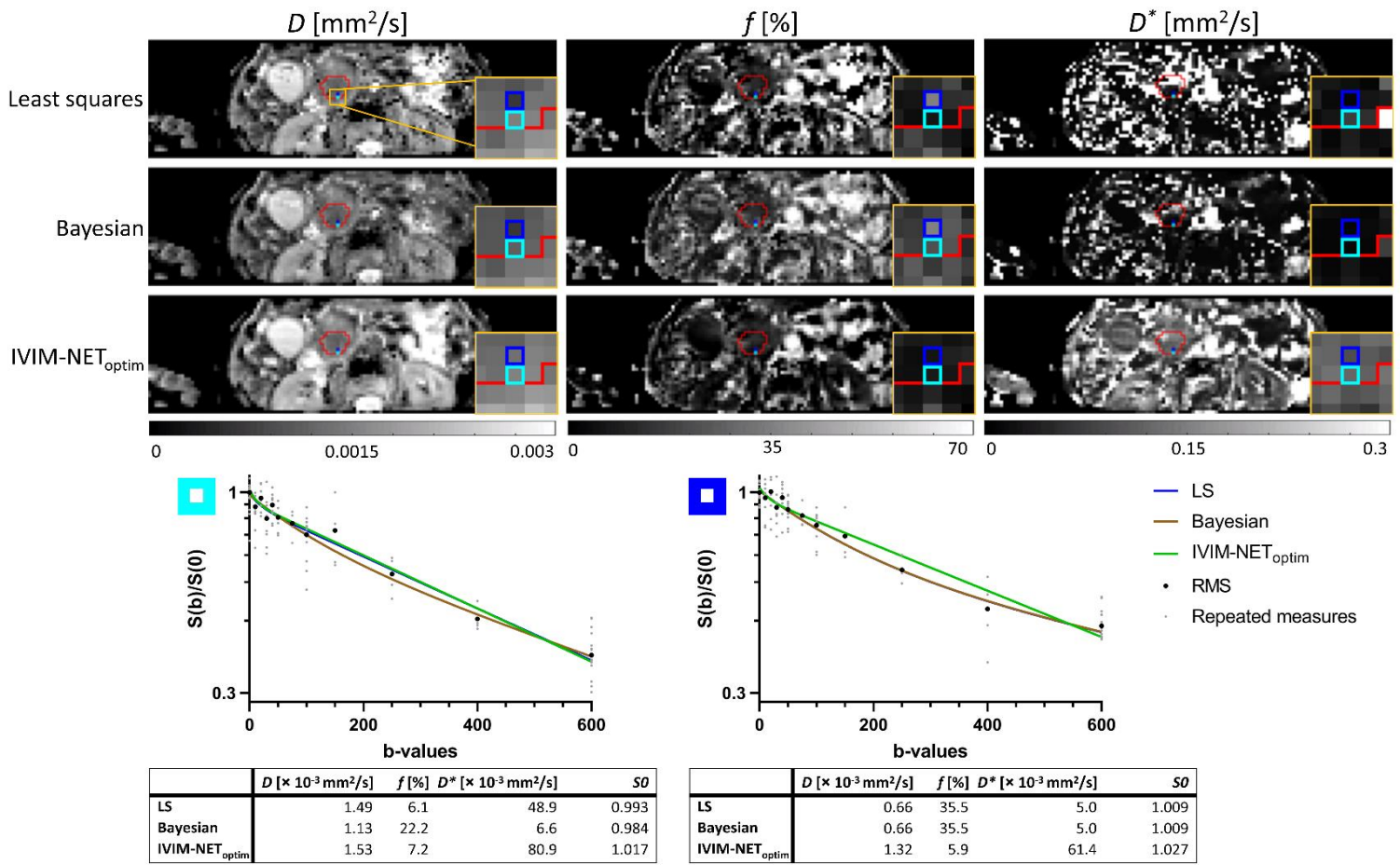

**Figure S15:** See Table S1.

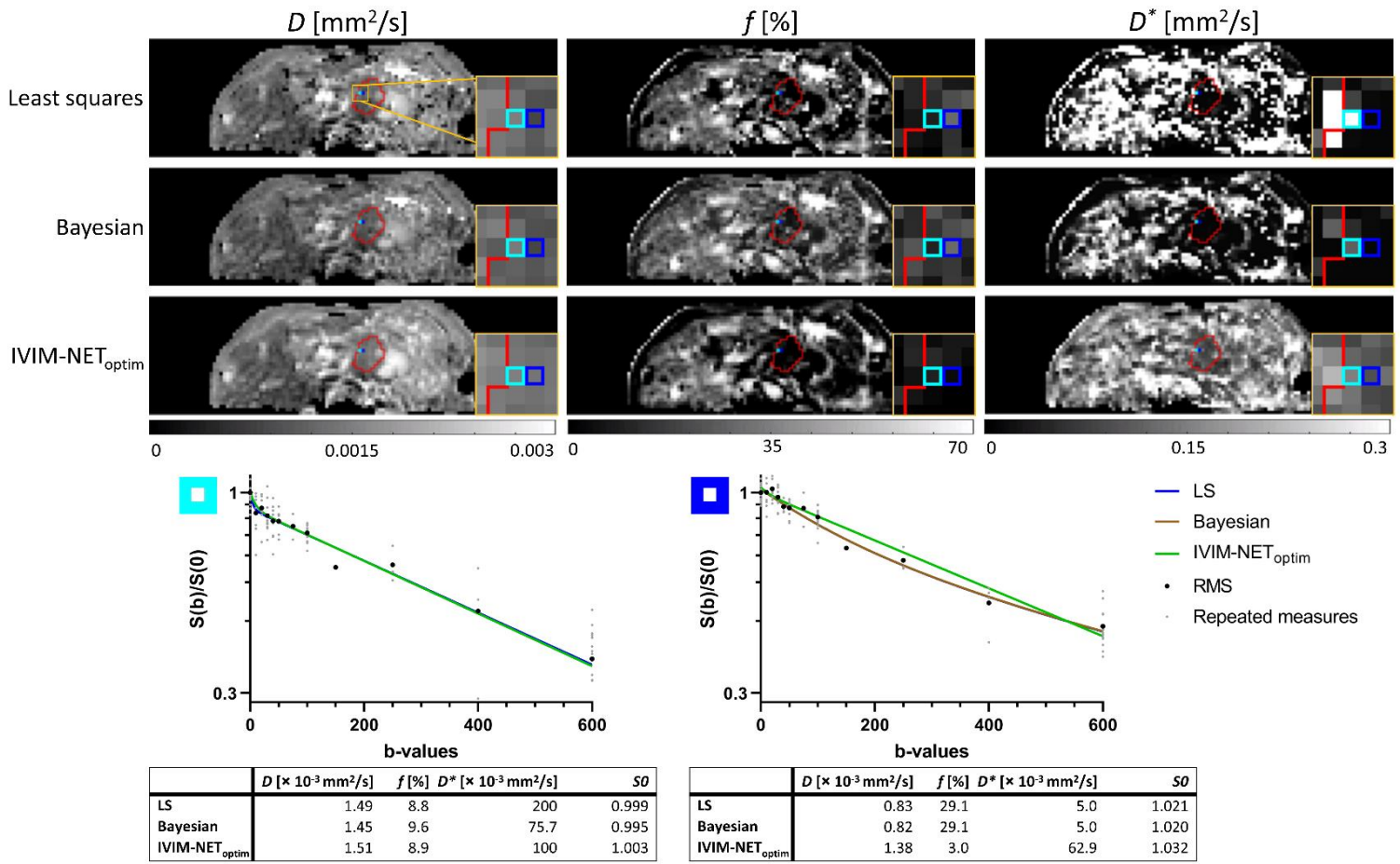

**Figure S16:** See Table S1.

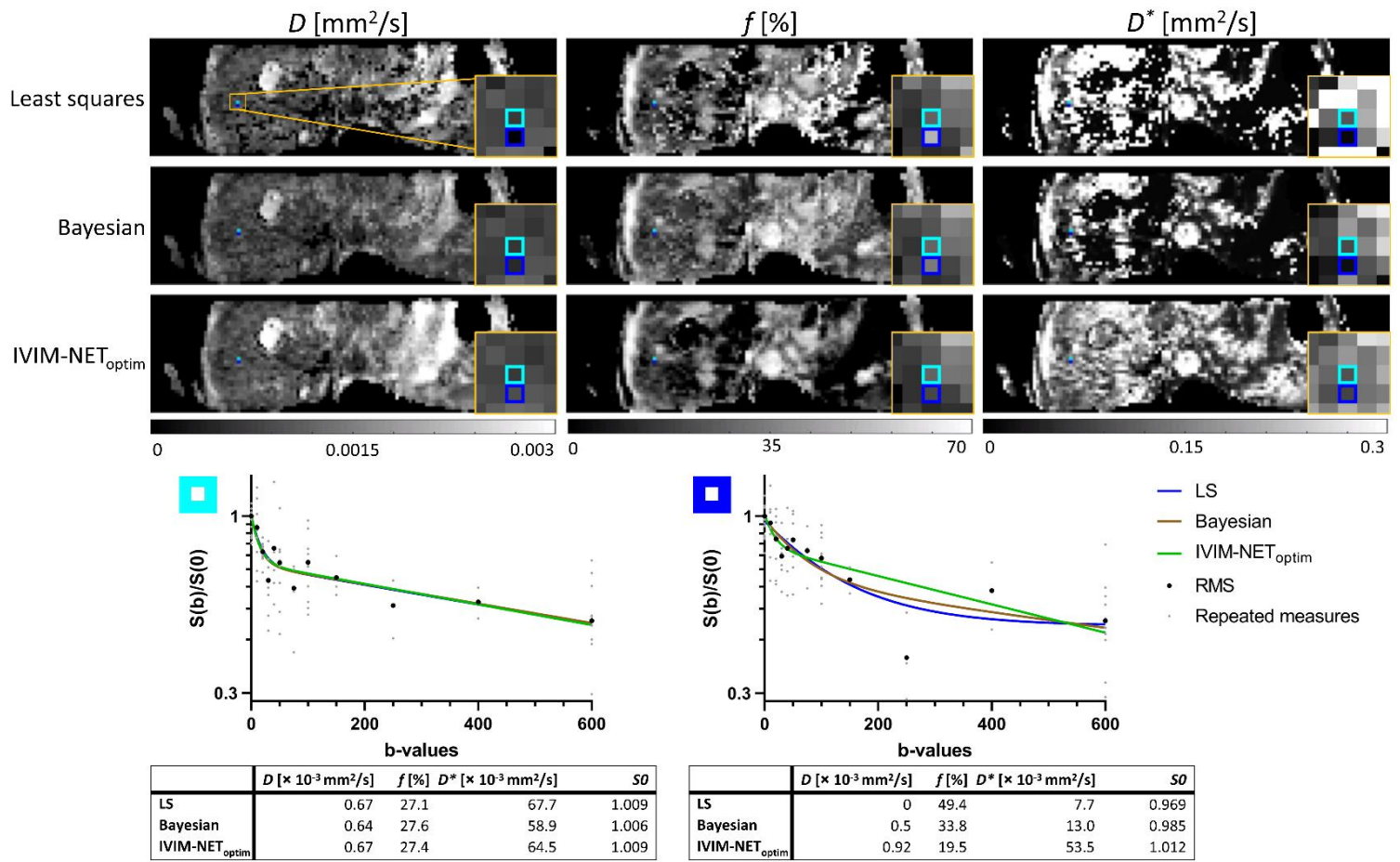

**Figure S17:** See table S1.

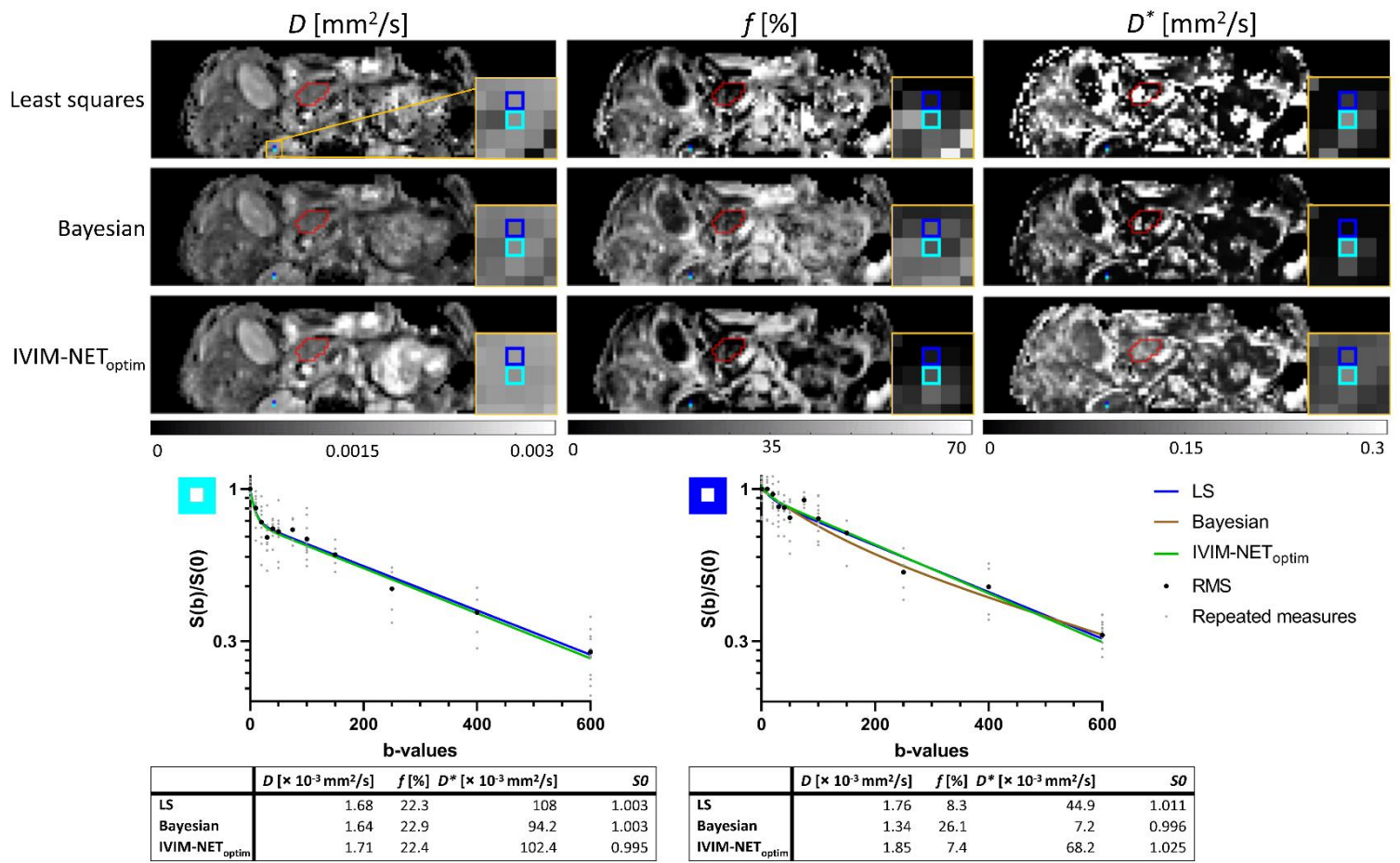

**Figure S18:** See Table S1.

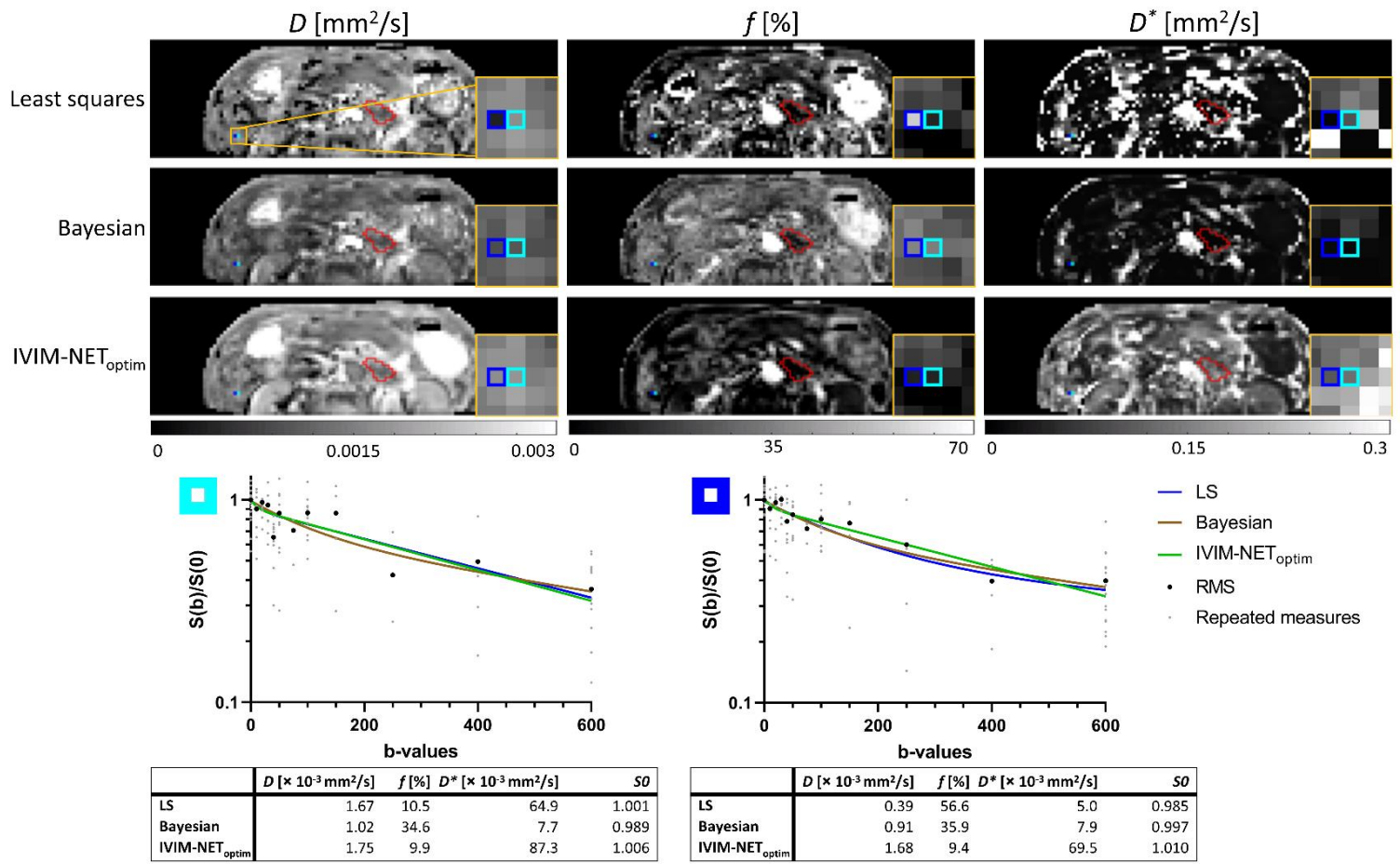

**Figure S19:** See Table S1.

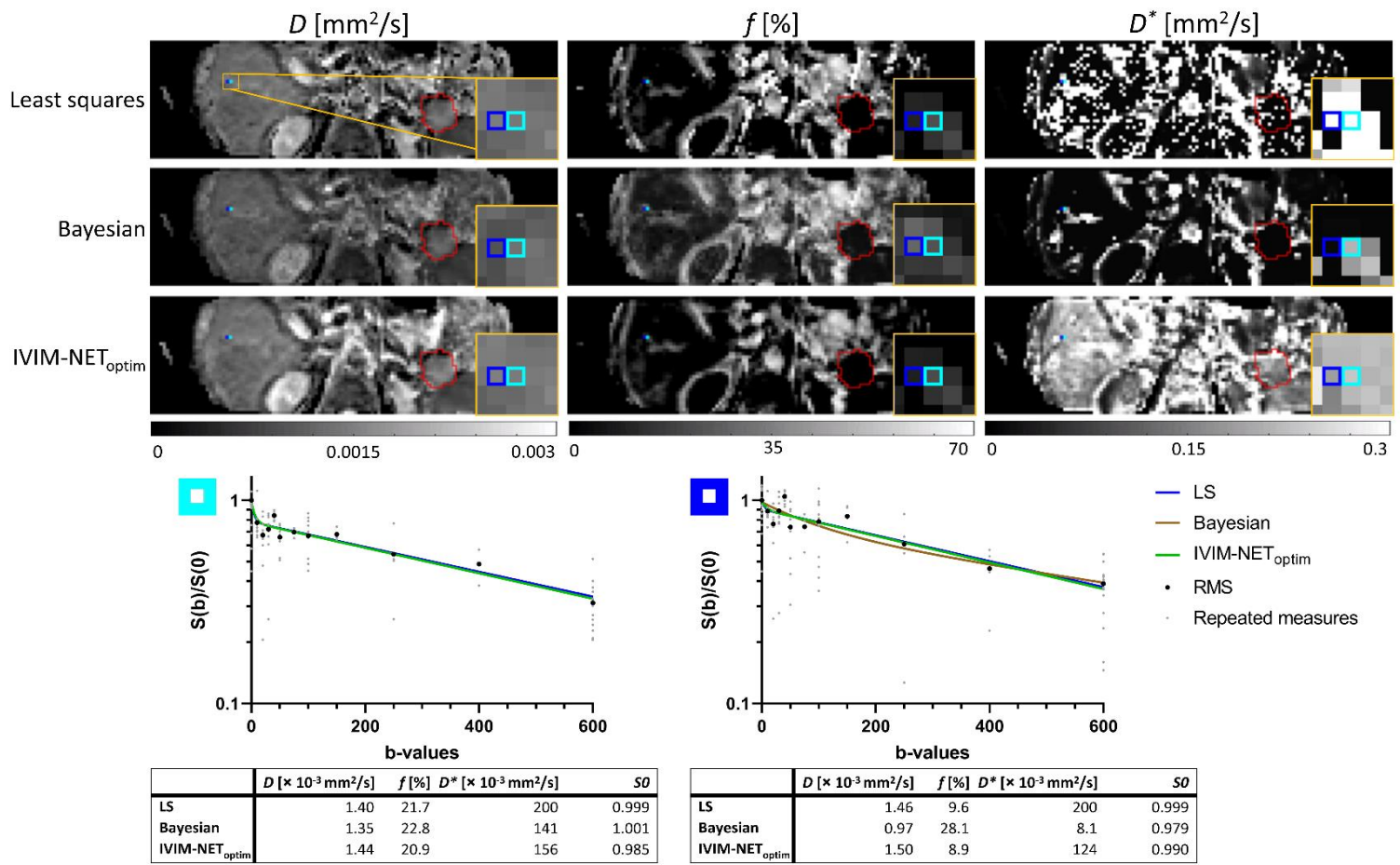

**Figure S20:** See Table S1.